\documentclass[twocolumn]{svjour3}       

\usepackage{latexsym}
\usepackage{bbm,epsf,graphicx,epsfig,psfrag,bm}
\usepackage{mathrsfs,amsmath,amssymb,textcomp}
\usepackage{tikz}
\usetikzlibrary{matrix,arrows}

\newcommand{\pf}[1]{\frac{\partial}{\partial #1}}
\newcommand{\pfq}[1]{\frac{\partial^{2}}{\partial^{2} #1}}

 \journalname{Journal of Computational Electronics}
\begin{document}

\title{Theory and simulation of quantum photovoltaic devices based on the non-equilibrium Green's
function formalism}

\titlerunning{NEGF theory of quantum photovoltaic devices}

\author{U. Aeberhard }


\institute{U. Aeberhard \at
              IEK-5 Photovoltaik \\
              Forschungszentrum J\"ulich\\
              D-52425 J\"ulich, Germany\\
              Tel.: +49 2461 61-2615\\
              Fax: +49 2461 61-2615\\
              \email{u.aeberhard@fz-juelich.de}           
}

\date{Received: date / Accepted: date}

\maketitle

\begin{abstract}
This article reviews the application of the non-equilibrium Green's function formalism to the
simulation of novel photovoltaic devices utilizing quantum confinement effects in low dimensional
absorber structures. It covers well-known aspects of the fundamental NEGF theory for a system of interacting electrons,
photons and phonons with relevance for the simulation of optoelectronic devices and introduces at the same time new
approaches to the theoretical description of the elementary processes of photovoltaic device operation, such as
photogeneration via coherent excitonic absorption, phonon-mediated indirect optical transitions or 
non-radiative recombination via defect states. While the description of the theoretical framework is
kept as general as possible, two specific prototypical quantum photovoltaic devices, a single
quantum well photodiode and a silicon-oxide based superlattice absorber, are used to illustrated the
kind of unique insight that numerical simulations based on the theory are able to provide.


\end{abstract}

\section{Introduction}
\label{intro}
The demand for higher photovoltaic energy conversion efficiencies has recently led to the emergence of a whole new
generation of solar cell concepts based on multiple junctions \cite{yamaguchi:05}, intermediate
bands \cite{luque:97}, multiple exciton generation \cite{nozik:05}	or hot carrier effects
\cite{ross:82}. Compared to the standard bulk technology, the corresponding devices are of highly
increased structural and conceptual complexity, comparable to the advanced optoelectronic devices in
the field of light emission, 	such as solid state lasers and light emitting diodes, and like these
counterparts, they strongly rely on the design 	degrees of freedom offered by the utilization of
semiconductor nanostructures.	 This instance is hardly surprising 	if one considers that
photovoltaics 	(``photons in - electrons out'') can be regarded as the inverse regime of 	solid
state lighting 	(``electrons in - photons out''). However, even though the structures be the same,
in switching 	to the photovoltaic operation, the focus is shifted from the optical to the electronic
output 	characteristics, with the objective 	to achieve maximum carrier extraction rate at largest
possible 	separation of contact Fermi levels.

 While there is still a relation between photovoltaic and light-emitting quantum efficiency through the optoelectronic
 reciprocity relation \cite{rau:07}, real devices are usually non-radiatively limited, and efficient charge transport
 from the location of photogeneration to and fast extraction at the contacts becomes thus of pivotal importance.
 Indeed, the semiconductor nanostructures not only have to provide tunable and localization enhanced optical
 transitions (photogeneration), but their absorbing states need either be coupled to extended bulk-like states or 
 among themselves in order to enable electronic current flow (charge separation). The coupling can be provided via
 hybridization in situations where carriers can	 tunnel between localized and extended states, as in the case of
 (multi) quantum well diodes, or via inelastic scattering processes such as absorption and emission
 of phonons or low energy photons, respectively. The former mechanism is important both in devices
 relying on thermionic	 emission, such as multi-quantum-well solar cells \cite{ned:99_2}, as well as
 in superlattice absorber devices in the sequential tunneling transport regime \cite{green:00}. The
 latter mechanism is utilized in the intermediate band solar cell, where the carriers from the
 isolated intermediate absorber band are excited into the contacted continuum via an additional photon absorption process.
 
 Apart from providing coupling mechanisms, non-radiative inelastic scattering mechanisms are also responsible 
 for any non-radiative output power losses in the transport process, either via recombination of charge carriers or via
 energy dissipation due to charge carrier thermalization and relaxation. A substantial reduction of this energy
 dissipation forms the basis of the hot carrier solar cell concept, which on the other hand requires the use of energy selective
 contacts to extract the hot carriers. Again, the application of semiconductor nanostructures is being considered, e.g.
 resonant tunneling through single quantum dots \cite{berghoff:08}. This is an example of a
 purely electronic use of nanostructures in solar cells, independent of the optical properties. Another example would be the interband tunnel
 contact in multi-junction devices \cite{hermle:08}. The opposite situation of purely
 optical functionality concerns the vast field of light management in solar cells via photonic structures,
 which lies beyond the scope of this review.
 
 As a result of the large number of implications that are linked to the use of nanostructures in photovoltaic devices,
 it is very hard to say whether a certain structure is going to be beneficial or detrimental to the device characteristics,
 even under the idealized conditions of the radiative limit, assuming defect-free interfaces etc. In many devices, there
 is for instance the intrinsic conflict mentioned above between maximum localization needed for ideal absorption and
 maximum delocalization required for ideal transport. Indeed, at this stage, only a few of the high-efficiency concepts
 have been realized successfully, such as the III-V semiconductor material based multi-junction devices and 
 strain-balanced multi-quantum-well solar cells, the latter example demonstrating in an impressive way the feasibility of highly efficient nanostructure
 based devices	with a large number of interfaces. In order to be able to asses the potentials and capabilities of
 the various novel photovoltaic concepts, a realistic theoretical estimate of the specific device characteristics is
 thus highly desirable.
 
 The complexity of structure and physical mechanisms as well as the prominent role of dimensional and quantum effects
 characterizing the operation of these novel solar cell devices preclude the use of standard macroscopic bulk
 semiconductor transport theory conventionally used in photovoltaics, which is nothing else than the charge continuity
 equation for electrons and holes with a source term corresponding to the net interband generation rate and a
 drift-diffusion current with a longitudinal electric field obeying Poisson's equation. On the other hand, rate equation type quantum
 optical approaches used for solid state lasers or light-emitting diodes often lack a satisfactory description of dissipative
 charge transport and extraction/injection at contacts, while suitable quantum transport formalisms were   devised for
 unipolar devices and do thus normally not include the optical coupling. Hence, for a comprehensive theoretical
 description of photovoltaic devices based on nanostructured absorbers and/or conductors, a theory
 is required that treats on equal footing both quantum optics and  dissipative quantum transport.	These requirements can only be 
 satisfied on the level of quantum kinetics, such as provided by the Wigner-function, density-matrix and non-equilibrium
 Green's function formalisms.
 
Among the quantum-kinetic theories, the non-equilibrium Green's function (NEGF) method is the most versatile and
powerful tool to study non-equilibrium properties of nanostructures, since it is based on a quantum field theoretical
approach to non-equilibrium statistical mechanics \cite{martin:59,schwinger:61,kadanoff:62,keldysh:65,langreth:76}. In
the NEGF approach, the Green's functions (GFs) for the involved (quasi)particles (electrons, holes, phonons, photons,
excitons, plasmons, etc) 	are the model functions providing the physical quantities that characterize the system.
They correspond to the response of the system to external perturbations, the latter entering the equations of motion for the Green's
functions, the Dyson's equations, in the form of self-energies. The self-energies due to (weak)
interactions such as electron-phonon or electron-photon coupling are calculated using standard
diagrammatic or operator expansion techniques used in many-body perturbation theory \cite{fetter_walecka}. The coupling
to the environment represented by contacts, i.e. the application of open boundary conditions, is provided by a special
type of boundary self-energy.
 
Due to the generality of the method steming form its sound foundations, the NEGF formalism has found application in the
description of large number of different micro- and mesoscopic systems under non-equilibrium conditions, of which only a
few can be mentioned here. Apart from the application to actual non-equilibrium quantum transport phenomena comprising
ballistic transport and resonant tunneling in semiconductor multilayers and nanostructures of different dimensionality
(quantum wells \cite{lake:97}, 	wires \cite{luisier:06_2,jin:06_2} and dots \cite{henrickson:94}), metallic and
molecular conduction \cite{tian:98,xue:01,brandbyge:02,dicarlo:05,stokbro:05,rocha:06,thygesen:08}, phonon
mediated inelastic 	and thermal transport \cite{lake:92,frederiksen:04,pecchia:07,lu:07,wang:07}, Coulomb-blockade
\cite{groshev:91,chen:91_2} and	 Kondo-effect  \cite{hershfield:91,meir:93}, it is also used to describe
strongly non-equilibrium and interacting regimes in semiconductor quantum optics requiring a quantum kinetic approach 
\cite{binder:95,haug:96,pereira:98,hannewald:01,schaefer:02,haug:04}, with phenomena such as non-equilibrium
absorption, interband polarization, spontaneous emission and laser gain. The concept was first adapted to the simulation
of 	transport in open nanoscale devices on the example of tunneling in metal-insulator-metal junctions \cite{caroli:71},
and has	 in the following been applied to investigation and modelling of
MOSFET\cite{svizhenko:02,guo:02,ren:03,jin:06,martinez:07},	CNT-FET \cite{guo:05,pourfath:06}, resonant tunneling	
diodes \cite{kim:88,anda:91,lake:92,kim:95,lake:97,ogawa:99} and interband tunneling diodes
\cite{ogawa:00,rivas:01,rivas:03,luisier:10}, 	interband quantum well lasers \cite{pereira:98} and intraband 
quantum cascade lasers \cite{lee_prb:02,kubis:09}, as well as infrared photodetectors \cite{henrickson:02}, 
CNT-photodiodes \cite{stewart:04,stewart:05,guo:06} and quantum well LEDs \cite{steiger:iwce_09}. The theory was		
formulated	 both for continuum effective-mass \cite{haug:96,jin:06} an atomistic multiband tight-binding \cite{lake:97}
models	 of the electronic structure, and in the case of molecular conduction it is combined with ab-initio methods such
as 	density functional theory \cite{brandbyge:02,evers:04,stokbro:05}.

 This review now provides an overview of the general framework for the theoretical description of quantum photovoltaic
 (QPV) devices based on the NEGF formalism, as first introduced in \cite{ae:prb_08,ae:thesis} for
 the investigation of quantum well solar cells (QWSC). After the specification of the type of system under
 consideration, the non-equilibrium quantum statistical treatment of its different physical degrees of freedom (charge carriers,
 photons, phonons) is discussed, resulting in a general microscopic theory of photovoltaic devices. The
 unique kind of spectral information the theory is able to provide is then illustrated on the example of two prototypical QPV devices, a single quantum
 well photodiode and a silicon-oxide based superlattice absorber device.
 
\section{Theoretical framework} 

As in any device simulation, the ultimate goal of the present approach it to predict the response of a 
device to a given variation in the external conditions (bias, temperature, illumination,\ldots) and for given material
properties (electronic, phononic, photonic structure), via the solution of (coupled) dynamical equations for the degrees
of freedom of the device affected by the perturbation. In a formalized picture, the dynamical equations describe
the evolution of the states of the system and their occupation for given initial solution and boundary conditions.
Before turning to the dynamical equations, let us thus briefly discuss the characteristic set-up of the system to be
described.
\label{sec:1} 

\begin{figure}[t]
\begin{center} 
\includegraphics[width=7cm]{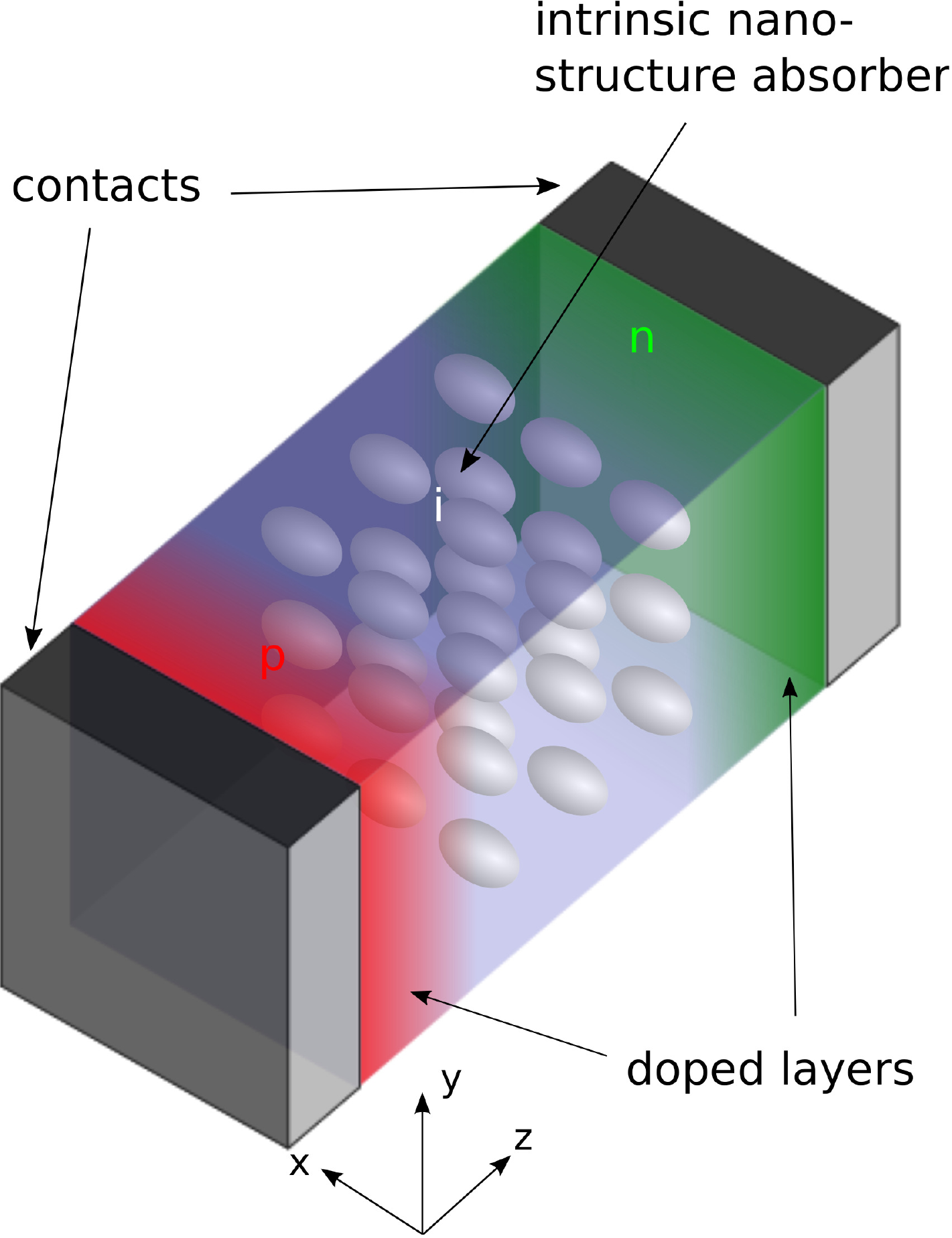}
\caption{Basic structure and functional elements of a generic quantum photovoltaic
device\label{fig:qpv_device_structure}.}
\end{center}
\end{figure}

\subsection{Definition of a QPV device}

\begin{figure}[t]  
\begin{center}
\includegraphics[width=8cm]{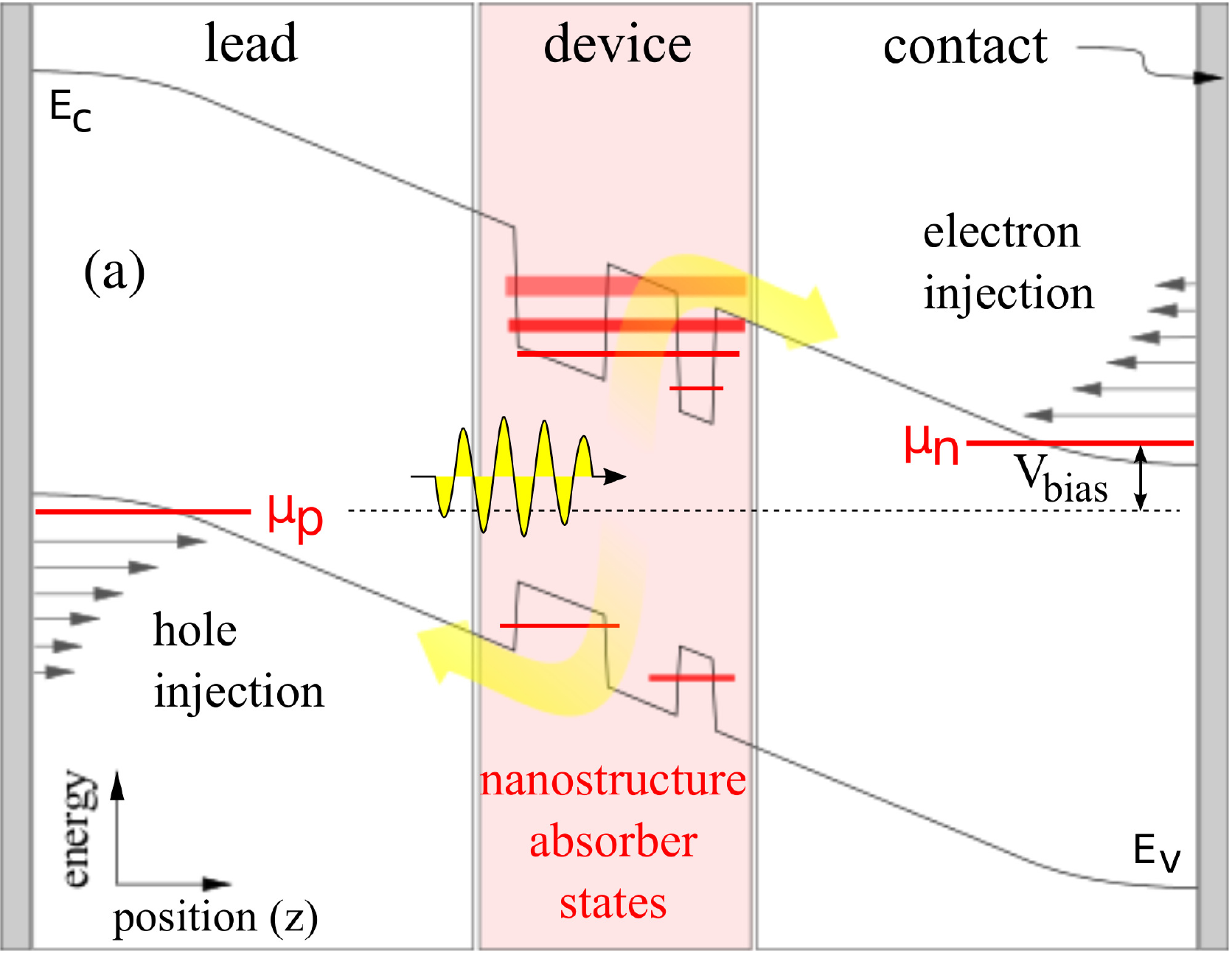} 
\end{center}
\begin{center}
\includegraphics[width=2.8cm]{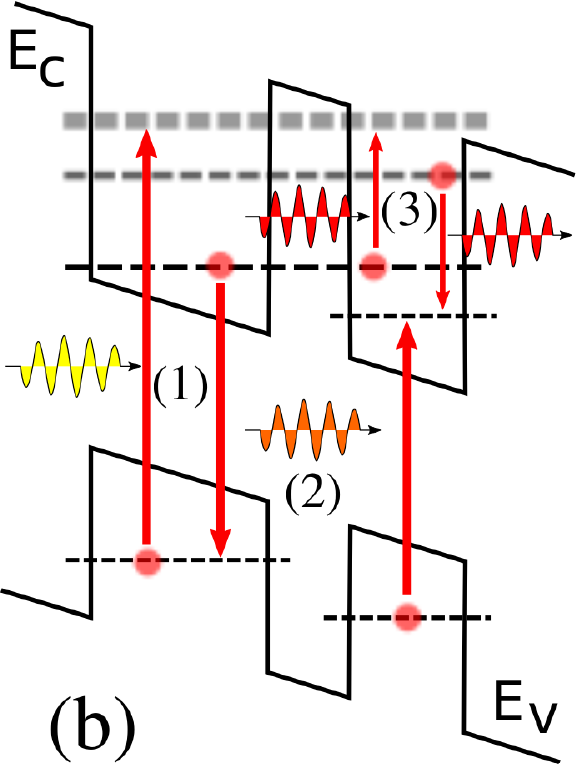}\includegraphics[width=2.8cm]{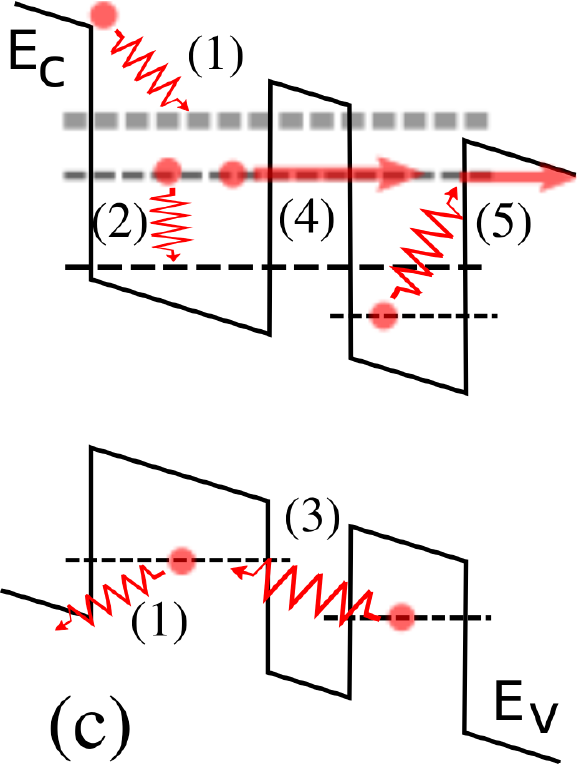}\includegraphics[width=2.8cm]{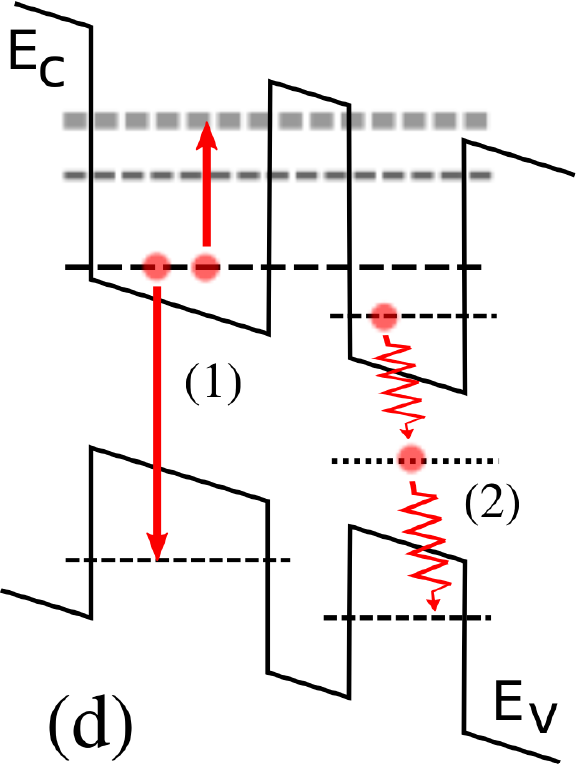}
\caption{(a) Energy band diagram corresponding to a device similar to the structure shown
in Fig. \ref{fig:qpv_device_structure}. The bulk leads are absent in the case where the nanostructure
states form both absorber and conductor. The device (absorber/emitter) states can be bound, quasi-bound or form a quasi-continuum.
At the contacts, photogenerated carriers are extracted and thermalized carriers are injected according to the chemical
potentials $\mu_{n,p}$ for electrons and holes, respectively. In the device regions, the physical processes relevant for
the photovoltaic operation are: (b) radiative transitions, i.e. (1) interband photogeneration and radiative recombination as well as (2) radiative intraband transitions, which may also lead to (3) photon recycling; (c) coherent and dissipative quantum transport involving non-radiative intraband transitions, such as (1) phonon
mediated carrier capture and escape, (2) intraband relaxation, (3) scattering assisted or (4) direct tunneling between	
absorber states, and (5) phonon assisted carrier escape as combination of the two; (d) non-radiative recombination, via
(1) the Auger mechanism or (2) deep trap states.
\label{fig:qpv_device_bandscheme}}
\end{center}
\end{figure}
Fig. \ref{fig:qpv_device_structure} shows the basic building blocks of a quantum photovoltaic
device, which are in essence  the same as for a conventional solar cell, namely an (intrinsic) absorber with carrier species selective contacts, usually obtained via
selective doping resulting in a bipolar junction device. The difference to the standard configuration is made by the
presence of additional nanostructure absorbers in the intrinsic region, leading to the formation of quantum wells, wires or dots and
thus exhibiting states with increased degree of localization. The band diagram along the
direction of transport which corresponds to such a device structure is shown schematically in
Fig. \ref{fig:qpv_device_bandscheme}(a). The bulk leads are absent in the case where the
nanostructure states form both absorber and conductor, i.e. in the case where the nanostructure states are extended in transport direction,
as in nanowires or superlattices. At the contacts, photogenerated carriers are extracted and
thermalized carriers are injected according to the splitting of the chemical potentials
$\mu_{n,p}$ of electron and hole reservoirs by the terminal voltage $V_{bias}$. Apart from this
functionality, 	which is common to all electronic devices, there is an aspect of contacts in
photovoltaics that is even more essential, which is the mentioned carrier selectivity. It can be
shown that even in absence of built in fields, the existence of carrier selective contacts enables
photovoltaic device operation \cite{wuerfel:book_05}. For some photovoltaic applications like the
hot carrier solar cell, contacts need in addition be energy selective. Even though most solar cells are
rather macroscopic than mesoscopic devices and the transport properties are not primarily
determined by the contact resistance, the contacts are usually critical components in the solar
cell design. This is due to electronic matching problems, leading to band offsets, Schottky
barriers, strong local fields and defect formation. Indeed, defect recombination at contacts
represents a major loss mechanism, and large efforts are dedicated to the passivation of these
surface	defects. While the contact states do normally form a continuum, the absorber states in the
device region can range from completely localized bound states to quasibound states or even extended 
quasicontinuum states, depending on the degree of confinement imposed by the nanostructure, and
they enable a large number of additional radiative and	 non-radiative transitions which are relevant for the photovoltaic device operation, as shown in
Figs. \ref{fig:qpv_device_bandscheme}(b)-(d). Here, the transitions which do involve only one
carrier species and conserve particle number are termed ''intraband'' processes, even though they take place between discrete states
rather within a proper band or even between subbands. Analogously, transitions across the
fundamental	gap ar termed ''interband'' transitions. In the case of absorber states with reduced
extension in transport direction, non-local processes such as direct or scattering assisted
tunneling connecting spatially separated states or localized to extended states become pivotal for the device
performance, and must thus be considered in any appropriate simulation of such a device.

\subsection{NEGF description of a QPV system}

Due to the large number of applications of the NEGF formalism, its foundations have been
discussed in many reviews and books, e.g. \cite{danielewicz:84,rammer:86,datta:95,schaefer:02},
wherefore we will limit the discussion here to the aspects that are relevant for photovoltaic applications. Many of
these aspects are also present in resonant-tunneling devices on the quantum transport side and in light-emitting devices on the quantum optics side,
and the discussion thus follows some of the lines of the corresponding literature, such as presented in \cite{lake:97}
and \cite{pereira:98}. We start the discussion with the quantum-statistical mechanics picture of the semiconductor
device.

\subsubsection{Field operators and Green's functions\label{sec:fieldop}}

A comprehensive, unified description of the quantum photovoltaic device is found on the quantum-statistical mechanics
level, where the optical, electronic and vibrational degrees of freedom of the system are described by the
corresponding quantum fields for photons, charge carriers and phonons.  Within the NEGF
theory of quantum optics and transport in excited  semiconductor nanostructures, physical quantities are expressed in
terms 	of quantum statistical ensemble averages of single particle operators for the interacting quasiparticles
introduced above, 	namely the fermion field operator $\hat{\Psi}$ for the charge carriers,  the quantized vector 
potential $\hat{\mathbf{A}}$ for the transverse photons and the ionic displacement field
$\hat{\boldsymbol{\mathcal{U}}}$ for the phonons.	The corresponding GFs are
\begin{align}
 G(\underbar{1},\underbar{2})=&-\frac{i}{\hbar}\langle\hat{\Psi}(\underbar{1})
 \hat{\Psi}^{\dagger}
 (\underbar{2})\rangle_{\mathcal{C}},~&\mathrm{(electrons)}\\
 \mathcal{D}_{\mu\nu}^{\gamma}(\underbar{1},\underbar{2})=&-\frac{1}{\mu_{0}}\frac{i}{\hbar}\Big[\langle
 \hat{A}_{\mu}(\underbar{1}) \hat{A}_{\nu}(\underbar{2})\rangle_{\mathcal{C}}\nonumber\\
&\qquad~-\langle A_{\mu}(\underbar{1})\rangle\langle A_{\nu}(\underbar{2})\rangle\Big]
~&\mathrm{(photons)}\label{eq:photgf} \\
  \mathcal{D}_{\alpha\beta}^{p}(\tilde{\underbar{1}},\tilde{\underbar{2}})=&-\frac{i}{\hbar}
  \langle\hat{\mathcal{U}}_{\alpha}
  (\tilde{\underbar{1}})\hat{\mathcal{U}}_{\beta}(\tilde{\underbar{2}})
  \rangle_{\mathcal{C}},~&\mathrm{(phonons)}  \label{eq:phongf}
 \end{align}   
where $\langle...\rangle_{C}$ denotes the contour ordered operator average 
peculiar to non-equilibrium quantum statistical mechanics
\cite{kadanoff:62,keldysh:65} for arguments
$\underbar{1}=(\mathbf{r}_{1},\underbar{t}_{1})$ with $\underbar{t}_{1}$ on the Keldysh
contour \cite{keldysh:65}. In the case of the lattice displacement field, the continuous spatial
coordinate $\mathbf{r}$ is replaced by the discrete ion position vectors
$\mathbf{R}=\mathbf{L}+\boldsymbol{\kappa}\equiv \mathbf{L}\boldsymbol{\kappa}$, where $\mathbf{L}$
is the equilibrium lattice vector and $\boldsymbol{\kappa}$ the basis vector within the unit cell
\cite{schaefer:02}. The associated four-vector is
$\tilde{\underbar{1}}\equiv(\mathbf{L}_{1}\boldsymbol{\kappa}_{1},\underbar{t}_{1})$.

Since it is a declared goal of the present approach to treat systems where any of the degrees
of freedom may be quantized to some extent via dimensional confinement, we will keep the treatment
as general as possible and thus use the real-space, complex-time
representation of the operators and their averages in the review of the formalism, even though in
many cases the choice of a reciprocal or modal space reduces formal and computational complexity
significantly. 

\subsubsection{Dynamical equations}

The GFs are determined as the solutions
of corresponding Dyson's equations \cite{henneberger:88_3,pereira:98,schaefer:02},
\begin{align}
\int_{\mathcal{C}}
d\underbar{3}\left[G_{0}^{-1}(\underbar{1},\underbar{3})-\Sigma(\underbar{1},\underbar{3})\right]
G(\underbar{3},\underbar{2}) &=\delta(\underbar{1}-\underbar{2}),\nonumber\\
\int_{\mathcal{C}} d\underbar{3}
\left[(\overleftrightarrow{\boldsymbol{\mathcal{D}}}_{0}^{\gamma})^{-1}(\underbar{1},\underbar{3})
- \overleftrightarrow{\boldsymbol{\Pi}}^{\gamma}(\underbar{1},\underbar{3})\right]
 \overleftrightarrow{\boldsymbol{\mathcal{D}}}^{\gamma}(\underbar{3},\underbar{2})&=
 \overleftrightarrow{\boldsymbol{\delta}}(\underbar{1}-\underbar{2}),\nonumber
 \label{eq:dyson_phot}\\ \int_{\mathcal{C}} d\underbar{3}
\left[(\boldsymbol{\mathcal{D}}_{0}^{p})^{-1}(\tilde{\underbar{1}},\tilde{\underbar{3}})-
\boldsymbol{\Pi}^{p}(\tilde{\underbar{1}},\tilde{\underbar{3}})\right]
\boldsymbol{\mathcal{D}}^{p}(\tilde{\underbar{3}},\tilde{\underbar{2}})&=
\boldsymbol{\delta}(\tilde{\underbar{1}}-\tilde{\underbar{2}}),
 \end{align}
 where the integration is $\int_{\mathcal{C}}d\underbar{1}\equiv\int_{\mathcal{C}}d t_{1}\int d^3
 r_{1}$. The GFs $G_{0}$, $\mathcal{D}_{0}^{\gamma}$ and $\mathcal{D}_{0}^{p}$ are the
propagators for non-interacting electrons, photons and phonons, respectively,  $\leftrightarrow$ denotes
transverse and boldface tensorial quantities. The electronic self-energy $\Sigma$ encodes the
renormalization of the charge carrier GFs due to the interactions with photons and
phonons and other carriers, and enables thus the description of charge carrier generation,
recombination and relaxation processes. It is responsible for the appearance of excitonic
effects in the carrier spectrum and  leads to band-gap renormalization under high excitation.
Charge injection and extraction at contacts is considered via an additional boundary self-energy 
term reflecting the openness of the system. The photon and phonon self-energy tensors 
$\overleftrightarrow{\boldsymbol{\Pi}}^{\gamma}$ and $\boldsymbol{\Pi}^{p}$ describe the
renormalization of the optical and vibrational excitation modes due to the interaction with the	
electronic system, i.e. absorption and emission of photons and phonons, leading to phenomena such 
as photon recycling or hot carrier effects and including excitonic signatures in the bosonic
spectra.

The self-energies can be derived either via perturbative methods using a diagrammatic approach or a
Wick factorization, or using variational derivatives. Their dependence on the GFs can
be expressed using Hedin's approach \cite{hedin:65} (henceforth, contour integration over internal variables is assumed)
\begin{align}
\Sigma(\underbar{1},\underbar{2})=&-i\hbar\mathcal{Q}
G(\underbar{1},\underbar{3})
\mathrm{\Gamma}^{e}(\underbar{3},\underbar{2},\underbar{4})\mathcal{W}_{eff}(\underbar{4},\underbar{1})\nonumber\\
&-i\hbar\vec{\mathcal{J}}(\underbar{1},\underbar{1}')G(\underbar{1},\underbar{3})\vec{\Gamma}^{\gamma}
(\underbar{3},\underbar{2},\underbar{4})\overleftrightarrow{\mathcal{D}}(\underbar{4},
\underbar{1}')|_{1=1'},\\
\overleftrightarrow{\boldsymbol{\Pi}}^{\gamma}(\underbar{1},\underbar{2})	
=&i\hbar\vec{\mathcal{J}}(\underbar{1}, \underbar{1}') G(\underbar{1},\underbar{3})
\vec{\Gamma}^{\gamma}(\underbar{3},\underbar{4},\underbar{2})
G(\underbar{4},\underbar{1'})|_{1=1'},\\ 
\boldsymbol{\Pi}^{p}(\tilde{\underbar{1}},\tilde{\underbar{2}})
=&i\hbar\vec{\mathcal{F}}(\tilde{\underbar{1}},\underbar{1}) G(\underbar{1},\underbar{3})
\vec{\Gamma}^{p}(\underbar{3},\underbar{4},\tilde{\underbar{2}})
G(\underbar{4},\underbar{1}),
\end{align}
where $\mathcal{Q}$, $\vec{\mathcal{J}}$ and $\vec{\mathcal{F}}$ are the bare
interaction vertices. In the expression for the phonon self-energy, the RHS is integrated over
all continuous spatial coordinates. The effective carrier-carrier interaction potential
$\mathcal{W}_{eff}$ is the sum of an electronic plasmon screening and a phonon-mediated
contribution,
\begin{align}
\mathcal{W}_{eff}(\underbar{1},\underbar{2})&=\mathcal{W}_{e}(\underbar{1},\underbar{2})+
\mathcal{W}_{p}(\underbar{1},\underbar{2}),\\ \mathcal{W}_{e}(\underbar{1},\underbar{2})&=
\mathcal{W}_{e(0)}(\underbar{1},\underbar{2})+\mathcal{W}_{e(0))}(\underbar{1},\underbar{4}) 
\pi(\underbar{4},\underbar{3})\mathcal{W}_{e}(\underbar{3},\underbar{2}),\\ 
\mathcal{W}_{p}(\underbar{1},\underbar{2})&=\mathcal{W}_{e}(\underbar{1},\underbar{4}) 
\mathcal{D}^{n}(\underbar{3},\underbar{4})\mathcal{W}_{e}(\underbar{4},\underbar{2}).
\end{align}
Here, $\pi$ is the longitudinal electromagnetic polarization function corresponding to the
self-energy of longitudinal photons and obeying the equation 
\begin{align}
\pi(\underbar{1},\underbar{2})&=i\hbar~
\mathcal{Q}G(\underbar{1},\underbar{3})\mathrm{\Gamma}^{e}(\underbar{3},\underbar{4},\underbar{2})
G(\underbar{4},\underbar{1})
\end{align}
and $\mathcal{D}^{n}$ is the density-density correlation function of the nuclei, which is related to
the phonon GF in real space via \cite{schaefer:02}
\begin{align}
&\mathcal{D}^{n}(\mathbf{r},t,\mathbf{r}',t')=\sum_{\mathbf{L}\boldsymbol{\kappa}\alpha}\sum_{\mathbf{L}'\boldsymbol{\kappa}'\beta}
\nabla_{\alpha}\rho_{\boldsymbol{\kappa}}(\mathbf{r}-\mathbf{L}-\boldsymbol{\kappa})\nonumber\\&\qquad\times
\mathcal{D}^{p}_{\alpha\beta}(\mathbf{L}\boldsymbol{\kappa},\mathbf{L}'\boldsymbol{\kappa}',t,t')
\nabla_{\beta}\rho_{\boldsymbol{\kappa}'}(\mathbf{r}'-\mathbf{L}'-\boldsymbol{\kappa}'),
\end{align}
where $\rho_{\boldsymbol{\kappa}}$ is the local charge density due to the ion cores.

The equations for the vertex corrections used in the iterative determination of the
self-energies are
\begin{align}
\mathrm{\Gamma}^{e}(\underbar{1},\underbar{2},\underbar{3})=&-\mathcal{Q}\delta(\underbar{1},\underbar{2})
\delta(\underbar{1},\underbar{3})+\frac{\delta\Sigma_{e\alpha}(\underbar{1},\underbar{2})}{\delta G(\underbar{4},
\underbar{5})}G(\underbar{4},\underbar{6})\nonumber\\
&\times\vec{\Gamma}^{\alpha}(\underbar{6}, \underbar{7},\underbar{3})
G(\underbar{7},\underbar{5}),\\
\vec{\Gamma}^{\gamma}(\underbar{1},\underbar{2},\underbar{3})=&-\mu_{0}\vec{\mathcal{J}}(\underbar{1},\underbar{1}')
\delta(\underbar{1},\underbar{2}) \delta(\underbar{1},\underbar{3})|_{1'=1}\nonumber\\ &
+\frac{\delta\Sigma(\underbar{1},\underbar{2})}{\delta G(\underbar{4},\underbar{5})}
G(\underbar{4},\underbar{6})\vec{\Gamma}^{\gamma}(\underbar{6},\underbar{7},\underbar{3})
G(\underbar{7},\underbar{5}),\\ \vec{\Gamma}^{p}(\underbar{1},\underbar{2},\underbar{3})
=&-\vec{\mathcal{F}}(\underbar{1}) \delta(\underbar{1},\underbar{2})
\delta(\underbar{1},\underbar{3})+\frac{\delta\Sigma(\underbar{1},\underbar{2})}{\delta
G(\underbar{4},\underbar{5})}G(\underbar{4},\underbar{6})\nonumber\\
&\times\vec{\Gamma}^{p}(\underbar{6},\underbar{7},\underbar{3})
G(\underbar{7},\underbar{5}).
\end{align}
Inserting these expressions in the equations for the polarization functions provides self-consistent
approximations for the latter in the form of the Bethe-Salpeter equation for the scalar four-point
polarization function
\begin{align}
\mathcal{P}(\underbar{1},\underbar{1}',\underbar{2},\underbar{2}')=&\mathcal{P}_{0}(\underbar{1},\underbar{1}',
\underbar{2},\underbar{2}')+\mathcal{P}_{0}(\underbar{1},\underbar{3},
\underbar{4},\underbar{1}')\nonumber\\&\times\mathcal{K}(\underbar{3},\underbar{4},\underbar{5},\underbar{6})
\mathcal{P}(\underbar{5},\underbar{6},\underbar{2},\underbar{2}'),\\
\mathcal{K}(\underbar{3},\underbar{4},\underbar{5},\underbar{6})=&\frac{\delta\Sigma(\underbar{3},\underbar{4})}{\delta
G(\underbar{5},\underbar{6})},
\end{align}
through the relations
\begin{align}
\pi(\underbar{1},\underbar{2})=&i\hbar\mathcal{Q}^2\mathcal{P}(\underbar{1},\underbar{1},\underbar{2},\underbar{2}),\\
\overleftrightarrow{\boldsymbol{\Pi}}^{\gamma}(\underbar{1},\underbar{2})=&i\hbar\vec{\mathcal{J}}(\underbar{1},\underbar{3})
\vec{\mathcal{J}}(\underbar{4},\underbar{4}')\mathcal{P}(\underbar{1},\underbar{3},\underbar{2},\underbar{4})\big|_{{4'=4
\atop 3=1} \atop 4=2},\\
\boldsymbol{\Pi}^{p}(\tilde{\underbar{1}},\tilde{\underbar{2}})=&i\hbar\vec{\mathcal{F}}(\tilde{\underbar{1}},\underbar{1}) \vec{\mathcal{F}}(\tilde{\underbar{2}},\underbar{2})\mathcal{P}(\underbar{1},\underbar{1},\underbar{2},\underbar{2}).
\end{align}

For practical evaluation, the above equations are rewritten in terms of the real time components of
the GFs and self-energies according to the order of the time arguments on the contour
\cite{keldysh:65,langreth:76}, i.e. the lesser ($<$), greater ($>$), retarded (R) and advanced (A)
components. In photovoltaics, in contrast to standard applications of NEGF in ultra-fast laser
physics, but similar to many other quantum transport situations, one is primarily interested in the
\emph{steady-state} behaviour\footnote{There are photovoltaic devices where the investigation of
time-dependent processes is crucial, e.g. the ultra-fast injection of charge from the dye into the
semiconductor in dye-sensitized solar cells.}, and it is thus usually sufficient to consider the
dependence on the time difference $\tau=t_{1}-t_{2}$, which permits a translation of the
investigation into the energy domain via the Fourier transform
\begin{align}
&O^{\alpha}({\mathbf r_{1}},{\mathbf
r}_{2};E)\equiv\int_{-\infty}^{\infty}d\tau
e^{\frac{i}{\hbar}E\tau}O^{\alpha}({\mathbf
r_{1}},{\mathbf r}_{2};\tau),\label{eq:fourttoe}\\
&O\in\{(G,D^{\gamma,p}),(\Sigma,\Pi^{\gamma,p})\},\quad\alpha\in\{R,A,<,>\}\nonumber,
\end{align}

\subsection{Hamiltonian and self-energies}

\subsubsection{System partitioning}

In the formulation of the system Hamiltonian, a common approach is to partition the system into easily accessible term,
with most of the complexity in the remainder. In the present case, there are three aspects of complexity that may be
tackled by this way, namely 1. the openness of the system, i.e. the requirement to include the contacts in the
description of the system, 2. the presence of interactions in the device, and 3. the different degrees of freedoms to
be considered, i.e. electronic, optical and vibrational. The first aspect leads to a system Hamiltonian partitioning of
the form
\begin{align}
\hat{H}=&\hat{H}^{D}+\hat{H}^{R}+\hat{H}^{DR}\label{eq:hampart_cont}
\end{align}
consisting of the separate terms for isolated device, contact reservoir and the coupling between
the two. In the second case, the system Hamiltonian is written as the sum of non-interacting and interaction terms,
\begin{align}
\hat{H}=&\hat{H}_{0}+\hat{H}_{i}.
\end{align}
Finally, we have the different components attributed to electrons, photons and phonons,
\begin{align}
\hat{H}=&\hat{H}_{e}+\hat{H}_{\gamma}+\hat{H}_{p},
\end{align}
where the terms include the mutual interactions as well as the interactions with further system components such as
impurities etc.

 In order to make the system treatable, interactions in the contacts are not considered explicitly, and
their effect on the contact-device coupling is neglected, which amounts to the assumption
$\hat{H}^{R}_{i}=\hat{H}^{DR}_{i}=0$. 	Furthermore, the bosonic subsystems will be regarded as
non-interacting, i.e. $\hat{H}_{i,\gamma\gamma}=\hat{H}_{i,pp}=0$. This leaves the total system
Hamiltonian
\begin{align}
\hat{H}=&\hat{H}_{0}+\hat{H}_{i}^{D},\\
\hat{H}_{0}=&\sum_{\alpha=e,\gamma,p}\left[\hat{H}^{D}_{0\alpha}+\hat{H}^{R}_{0\alpha}
+\hat{H}^{DR}_{0\alpha}\right],\\
\hat{H}_{i}^{D}=&\hat{H}_{ee}^{D}+\hat{H}_{e\gamma}^{D}+\hat{H}_{ep}^{D}.
\end{align}

To quantify the action of the Hamiltonian on the many-body system, it is represented in second quantization using the
field operators introduced in Sec.\ref{sec:fieldop}, e.g. for the electronic system
\begin{align}
\mathcal{H}(t)&=\int d^{3}r 
\hat{\Psi}^{\dagger}(\mathbf{r},t)\hat{H}_{e}\hat{\Psi}(\mathbf{r},t).
\end{align}
In the case of the carrier-carrier interaction, this produces a two-particle Hamiltonian, while all the other terms are
on the single-particle level.

The Hamiltonian can now be used in the derivation of the equations for the self-energies using the
many-body perturbation theoretical expansion of
\begin{align}
G(\mathbf{r},t;\mathbf{r}',t')= -\frac{i}{\hbar}\left\langle
e^{-\frac{i}{\hbar}\int_{C}ds
\mathcal{\hat{H}}'(s)}\hat{\Psi}(\mathbf{r},t)\hat{\Psi}^{\dagger}(\mathbf{r}',t')\right\rangle_{C},\label{eq:pertexp_el}
\end{align}
and 
\begin{align}
&\mathcal{D}^{\alpha}_{\mu\nu}(\mathbf{r},t;\mathbf{r}',t')=
\frac{-if_{\alpha}}{\hbar}\Big\langle e^{-\frac{i}{\hbar}\int_{C}ds
\mathcal{\hat{H}}'(s)}\hat{\mathcal{F}}_{\mu}^{\alpha}(\mathbf{r},t)\hat{\mathcal{F}}_{\nu}^{\alpha\dagger}
(\mathbf{r}',t')\Big\rangle_{C},\nonumber\\
&\qquad\alpha=(\gamma,p),~
f_{\gamma,p}=(\mu_{0}^{-1},1),~\mathcal{F}^{\gamma,p}=(A,\mathcal{U}),\label{eq:pertexp_bos}
\end{align}
 where the perturbation is  given by  $\mathcal{\hat{H}}'=\hat{\mathcal{H}}-\hat{\mathcal{H}}_{0}^{D}$. By
 restricting the perturbation to specific parts of the Hamiltonian, the corresponding self-energy terms can be identified in the
 resulting Dyson equation. In the following, the Hamiltonian terms and corresponding self-energies for the contacts and
 the various interactions shall be discussed.
 
\subsubsection{Non-interacting isolated subsystems}
 
\paragraph{Electrons}
 
The electronic system without interactions\footnote{In contrast to the case of photons and phonons,
there are no completely free electrons in the system, since there is always the lattice potential
due to the ion cores.} between valence electrons and coupling to the bosonic degrees of freedom is
described by
\begin{align}
\hat{H}_{e}^{(0)}(\mathbf{r})=&-\frac{\hbar^{2}}{2m_{0}}\Delta_{\mathbf{r}}+\tilde{U}(\mathbf{r}),\label{eq:nint_ham_el}
\end{align}
where $m_{0}$ is the free electron mass. In the above expression, the first term provides the
kinetic energy and $\tilde{U}$ contains the potential for the interaction of valence electrons with 
the ion cores. It is further customary to include in $\tilde{U}$ also the Hartree term of the
Coulomb interaction corresponding to the solution of Poisson's equation that considers
carrier-carrier	interactions ($\hat{H}_{ee}$) on a mean-field level. The GFs
corresponding to this Hamiltonian is defined via
\begin{equation}
\left[i\hbar\pf{t_{1}}-\hat{H}_{e}^{(0)}(\mathbf{r})\right]G_{0}(1,1')=\delta(1,1'), 
\end{equation}
with the inverse function in \eqref{eq:dyson_phot} given by
\begin{equation}
G_{0}^{-1}(1,1')=\left[i\hbar\pf{t_{1}}-\hat{H}_{e}^{(0)}(\mathbf{r})\right]\delta(1,1').
\label{eq:inv_carr}
\end{equation}

The Hamiltonian operator in \eqref{eq:nint_ham_el} is the one entering the steady-state
Schr\"odinger equation corresponding to the single particle eigenvalue problem of the isolated system,
\begin{align}
\hat{H}_{e}^{(0)}|\psi_{n}\rangle=&\varepsilon_{n}|\psi_{n}\rangle.
\end{align}
In photovoltaic devices, the electronic system is open in at least one dimension $(r')$, for which
thus scattering states have to be used rather than eigenstates of the closed system. However, the remaining confined
dimensions $(\tilde{\mathbf{r}})$ may still be described by the corresponding eigensolutions,
allowing for an eigenmode expansion of Hamiltonian, GFs and self-energies in these dimensions, which results in
a general field operator representation
\begin{align}
\hat{\Psi}(\mathbf{r},t)=\sum_{\tilde{n}}\varphi_{\tilde{n}}^{e}(\tilde{\mathbf{r}})\hat{\chi}_{\tilde{n}}(r'),\quad
\mathbf{r}=(\tilde{\mathbf{r}},r').
\end{align}

\paragraph{photons}

While absorption in most nanostructured solar cells may be reasonably described by the linear
response to a \emph{coherent} excitation, spontaneous emission as the most fundamental loss process
relies on the coupling to the \emph{incoherent} field due to vacuum fluctuations. In general, the
total electromagnetic field may thus be written as the superposition
$\hat{\mathbf{A}}=\hat{\mathbf{A}}_{coh}+\hat{\mathbf{A}}_{inc}$ of a coherent contribution
$\hat{\mathbf{A}}_{coh}\equiv\langle \hat{\mathbf{A}}\rangle$ corresponding to coherent light 
sources and a contribution $\hat{\mathbf{A}}_{inc}$ with  $\langle \hat{\mathbf{A}}_{inc}\rangle=0$ 
from incoherent light sources and from spontaneous emission.  While the coherent vector potential is related to the
time-dependent part of the classical electric field $\vec{\mathcal{E}}$ via the standard relation
\begin{align}
-\pf{t}\mathbf{A}_{coh}(\mathbf{r},t)=\vec{\mathcal{E}}(\mathbf{r},t),
\end{align}
and is thus obtained from the solutions of Maxwell's equations according to classical electrodynamics, the incoherent
part needs to be treated on the quantum mechanical level via the GF introduced above and will be discussed in
what follows.

In terms of the vector potential operator introduced at the beginning of this section, the Hamiltonian of transverse
electromagnetic radiation reads
\cite{schaefer:02}
\begin{align}
\mathcal{H}_{\gamma}^{(0)}(t)=\frac{\varepsilon_{0}}{2}\int
d^{3}\mathbf{r}\left[\left(\pf{t}\hat{\mathbf{A}}(\mathbf{r},t)\right)^{2}+c_{0}^{2}
\left(\nabla\times\hat{\mathbf{A}}(\mathbf{r},t)\right)^{2}\right].
\end{align}
The free contour-ordered photon GF in real space obeys the equation
\begin{equation}
\left[\Delta_{\mathbf{r}_{1}}-\frac{1}{c_{0}^{2}}\pfq{\underbar{t}_{1}}\right]
\overleftrightarrow{\boldsymbol{\mathcal{D}}}_{0}(1,1') =\overleftrightarrow{\boldsymbol{\delta}}(1,1').
\end{equation}
which, in analogy to \eqref{eq:inv_carr}, defines the inverse free photon GF
entering the photon Dyson equation in \eqref{eq:dyson_phot} via
\begin{align}
\overleftrightarrow{\boldsymbol{\mathcal{D}}}_{0}^{-1}(1,1')=\left[\Delta_{\mathbf{r}_{1}}
-\frac{1}{c_{0}^{2}}\pfq{\underbar{t}_{1}}\right]\overleftrightarrow{\boldsymbol{\delta}}(1,1').\label{eq:freephotprop}
\end{align}

Since it is a solution to the wave equation following from Maxwell's equations,
the quantized field can be represented as the free-field mode expansion 
\begin{align}
\hat{\mathbf{A}}({\mathbf r},t)=&\sum_{\lambda,{\mathbf
q}}\left[\mathbf{A}_{0}(\lambda,\mathbf{q}) \hat{a}_{\lambda,{\mathbf
q}}(t)+\mathbf{A}_{0}^{*}(\lambda,-\mathbf{q}) 
\hat{a}_{\lambda,{-\mathbf q}}^{\dagger}(t)\right]\nonumber\\
&\times e^{i{\mathbf q}{\mathbf
r}},\label{eq:photfieldop}\\
\mathbf{A}_{0}(\lambda,\mathbf{q})=&\frac{\hbar}{\sqrt{2\epsilon_{0}V\hbar\omega_{\mathbf{q}}}}
\boldsymbol{\epsilon}_{\lambda{\mathbf q}},
\end{align}
where ${\boldsymbol \epsilon}_{\lambda{\mathbf q}} $ is the polarization of the
photon with wave vector 
${\mathbf q}$ and energy $\hbar\omega_{\mathbf{q}}$ added to or removed from photon
mode $(\lambda,\mathbf{q})$ by the bosonic creation and annihilation operators 
\begin{align}
\hat{a}_{\lambda,{\mathbf q}}^{\dagger}(t)=&\hat{a}_{\lambda,{\mathbf
q}}^{\dagger}e^{i\omega_{\lambda,{\mathbf q}}t},\quad \hat{a}_{\lambda,{\mathbf
q}}(t)=\hat{a}_{\lambda,{\mathbf q}}e^{-i\omega_{\lambda,{\mathbf q}}t},\label{eq:timeev_photop}
\end{align}
and $V$ is the volume. In this representation, the non-interacting Hamiltonian of transverse
radiation acquires the simple form
\begin{equation}
\hat{\mathcal{H}}^{(0)}_{\gamma}(t)=\sum_{\lambda,{\mathbf
q}}\hbar\omega_{\lambda,\mathbf{q}}\left(\hat{a}_{\lambda,{\mathbf
q}}^{\dagger}(t)\hat{a}_{\lambda,{\mathbf q}}(t)+\frac{1}{2}\right).
\label{eq:photham}
\end{equation}
With the representation \eqref{eq:photfieldop}, the EM GF in \eqref{eq:photgf} can be
related to the bare scalar photon GF corresponding to the following expectation value of bosonic
operators,
\begin{align}
\mathcal{D}_{\lambda}^{\gamma,(0)}(\mathbf{q};\underbar{t},\underbar{t}')\equiv&
-\frac{i}{\hbar}\left\langle
\hat{a}_{\lambda,\mathbf{q}}(\underbar{t})\hat{a}^{\dagger}_{\lambda,\mathbf{q}}(\underbar{t}')
+\hat{a}^{\dagger}_{\lambda,-\mathbf{q}}(\underbar{t})\hat{a}_{\lambda,-\mathbf{q}}(\underbar{t}')\right\rangle_{\mathcal{C}}
\end{align} 
via 
\begin{align}
\mathcal{D}_{\mu\nu}^{\gamma,(0)}(\underbar{1},\underbar{1}')=&\sum_{\lambda,\mathbf{q}}
\frac{\hbar}{2\varepsilon_{0}V\omega_{\lambda,\mathbf{q}}}
e^{i\mathbf{q}\cdot(\mathbf{r}_{1}-\mathbf{r}_{1}')} \epsilon^{\mu}_{\lambda{\mathbf q}}\epsilon^{\nu}_{\lambda{\mathbf
q}}\nonumber\\&\times\mathcal{D}_{\lambda}^{\gamma,(0)}(\mathbf{q};\underbar{t}_{1},\underbar{t}_{1'}).
\end{align}
Noting that the photon occupation number operator for free field mode $(\lambda,\mathbf{q})$ is given by
$\hat{n}_{\lambda,\mathbf{q}}=\hat{a}^{\dagger}_{\lambda,\mathbf{q}}\hat{a}_{\lambda,\mathbf{q}}$,
the components of the scalar photon GF can be expressed in the energy domain as
\begin{align}
  \mathcal{D}_{\lambda}^{\gamma,\lessgtr}({\mathbf q};E) =&-2\pi
  i\Big[N^{\gamma}_{\lambda,{\mathbf q}}\delta(E\mp \hbar\omega_{\lambda,{\mathbf q}})\nonumber\\
  &+(N^{\gamma}_{\lambda,{\mathbf
        q}}+1)\delta(E\pm\hbar\omega_{\lambda,{\mathbf q}})\Big],\\
  \mathcal{D}_{\lambda}^{\gamma,R/A}({\mathbf q};E)&=\frac{1}{E-\hbar\omega_{\lambda,{\mathbf
  q}}\pm i\eta}-\frac{1}{E+\hbar\omega_{\lambda,{\mathbf q}}\pm i\eta}.
\label{eq:equilphoten}
\end{align}
with $N^{\gamma}_{\lambda,\mathbf{q}}=\langle \hat{n}^{\gamma}_{\lambda,\mathbf{q}}\rangle$ the  mode occupation, which
in the free-field case is related to the incident photon flux via \begin{align}
N^{\gamma}_{\lambda,\mathbf{q}}=\phi^{\gamma}_{\lambda,\mathbf{q}}V/c_{0}.
\end{align}
The modal photon flux in turn is given by the modal intensity of the EM field through
$\phi^{\gamma}_{\lambda,\mathbf{q}}=I^{\gamma}_{\lambda,\mathbf{q}}/(\hbar\omega_{\lambda,\mathbf{q}})$.
For normal incidence along $r'$  $(\hat{\vec{r}}'\perp \tilde{\vec{r}})$ and thermal radiation from source at
temperature $T$, the occupation can be written \cite{mozyrsky:07}
\begin{align}
N_{\lambda,\mathbf{q}}^{\gamma}=n_{BE}(\hbar c_{0}|\mathbf{q}|,T)\theta(q')\theta(q'\tan\alpha
-|\tilde{\mathbf{q}}|),
\end{align}
where $n_{BE}$ is the Bose-Einstein distribution function and $\alpha$ is half of the aperture angle under which the
source is seen by the solar cell.

In the general case of an inhomogeneous medium, one may opt to include the modification of the optical
mode due to spatially varying background dielectric constant $\varepsilon_{b}(\mathbf{r})$ in
\eqref{eq:freephotprop} for the ''free'' propagator by replacing $c_{0}$ by
$c(\mathbf{r})=c_{0}/n_{0}(\mathbf{r})$, where $n_{0}=\sqrt{\varepsilon_{b}}$. In the waveguide situation, the resulting problem gives rise to confined optical
modes, which, in analogy to the electronic case, can be used to expand the photon GFs \cite{jahnke:95}.
As discussed above, solar cells are optically open in at least one dimension $(r')$.  The general expression for the
photon field operator is thus 
\begin{align}
\hat{\mathbf{A}}(\mathbf{r},t)=\sum_{n}\varphi^{\gamma}_{n}(\tilde{\mathbf{r}})\hat{\mathbf{A}}_{n}(r',t),
\quad \mathbf{r}=(\tilde{\mathbf{r}},r'),
\end{align}
where $\varphi_{n}^{\gamma}(\mathbf{r})$ are the eigenmodes in the confined dimensions.

\paragraph{phonons}
  
In analogy to the photon field, the lattice displacement field for a bulk semiconductor with full translational
symmetry can be expanded into vibrational eigenmodes in reciprocal space,
\begin{align}
\hat{\mathcal{U}}_{\alpha}(\mathbf{L}\boldsymbol{\kappa},t)=&\sum_{\mathbf{Q}}
\sum_{\Lambda}\mathcal{U}_{0,\alpha\boldsymbol{\kappa}}(\Lambda,\mathbf{Q})
\big[\hat{b}_{\Lambda,\mathbf{Q}}(t)+\hat{b}^{\dagger}_{\Lambda,-\mathbf{Q}}(t)\big]\nonumber\\
&\times e^{i\mathbf{Q}\cdot(\mathbf{L}+\boldsymbol{\kappa})},\label{eq:normalmode}\\
\mathcal{U}_{0,\alpha\boldsymbol{\kappa}}(\Lambda,\mathbf{Q})=&
\frac{\hbar}{\sqrt{2NM_{\boldsymbol{\kappa}}\hbar\Omega_{\Lambda,\mathbf{Q}}}}
\boldsymbol{\epsilon}_{\alpha\boldsymbol{\kappa}\Lambda}({\mathbf Q}), \\
&(\alpha=x,y,z),\nonumber
\end{align}
where $N$ is the number of atoms, $M_{\boldsymbol{\kappa}}$ the mass of the basis atom at $\boldsymbol{\kappa}$, and
$\hat{b}_{\Lambda,{\mathbf Q}},\hat{b}_{\Lambda,{\mathbf Q}}^{\dagger}$ are the bosonic creation and annihilation
operators for a	 phonon mode $(\Lambda, {\mathbf Q})$ with polarization
$\boldsymbol{\epsilon}_{\alpha\boldsymbol{\kappa}	\Lambda}({\mathbf Q})$ and energy $\hbar\Omega_{\Lambda,\mathbf{Q}}$ 
determining the time-evolution in complete analogy to \eqref{eq:timeev_photop}. The phonon modes of a given structure follow from the eigenvalue equation
\begin{align}
\sum_{\boldsymbol{\kappa}',\beta}\mathcal{M}_{\alpha\boldsymbol{\kappa},\beta\boldsymbol{\kappa}'}(\mathbf{Q})
\boldsymbol{\epsilon}_{\alpha\boldsymbol{\kappa}\Lambda}({\mathbf Q})=&(\hbar\Omega_{\Lambda,\mathbf{Q}})^{2}
\boldsymbol{\epsilon}_{\alpha\boldsymbol{\kappa}\Lambda}({\mathbf Q}), \label{eq:phonon_evp}
\end{align}
in terms of the dynamical matrix $\mathcal{M}$, which in turn is determined by the interatomic force constants $\Phi$
via
\begin{align}
\mathcal{M}_{\alpha\boldsymbol{\kappa},\beta\boldsymbol{\kappa}'}(\mathbf{Q})
=&\frac{\hbar}{\sqrt{M_{\boldsymbol{\kappa}}M_{\boldsymbol{\kappa}}'}}\sum_{\mathbf{L}}
e^{-i\mathbf{Q}\cdot(\mathbf{L}+\boldsymbol{\kappa}-\boldsymbol{\kappa}')}
\Phi_{\alpha\beta}(\mathbf{L}\boldsymbol{\kappa},\mathbf{0}\boldsymbol{\kappa}).
\end{align}
The interatomic force constants can be obtained via phenomenological models or ab-initio calculations
\cite{srivastava:90,luisier:09}. 
 Again, it is possible to express the Hamiltonian and GF of the displacement field
 given in \eqref{eq:phongf} in terms of the vibrational eigenmodes, the former acquiring the form
 \begin{align}
 \hat{\mathcal{H}}^{(0)}_{p}(t)=\sum_{\Lambda,{\mathbf
Q}}\hbar\Omega_{\Lambda,\mathbf{Q}}\left(\hat{b}_{\Lambda,{\mathbf
Q}}^{\dagger}(t)\hat{b}_{\Lambda,{\mathbf Q}}(t)+\frac{1}{2}\right).
\label{eq:phonham}
 \end{align}
 and in the case of the latter via the bare scalar bosonic propagator 
 \begin{align}
\mathcal{D}_{\Lambda}^{p}(\mathbf{Q};\underbar{t},\underbar{t}')\equiv&
-\frac{i}{\hbar}\left\langle
\hat{b}_{\Lambda,\mathbf{Q}}(\underbar{t})\hat{b}^{\dagger}_{\Lambda,\mathbf{Q}}(\underbar{t}')
+\hat{b}^{\dagger}_{\Lambda,-\mathbf{Q}}(\underbar{t})\hat{b}_{\Lambda,-\mathbf{Q}}(\underbar{t}')\right\rangle_{\mathcal{C}},
 \end{align}
 through the relation
 \begin{align}
 \mathcal{D}_{\alpha\boldsymbol{\kappa},\beta\boldsymbol{\kappa'}}^{p}(\underbar{1},\underbar{1}')=&\sum_{\Lambda,\mathbf{Q}}
\frac{\hbar}{2\varepsilon_{0}N^2M_{\boldsymbol{\kappa}}M_{\boldsymbol{\kappa}'}\Omega_{\Lambda,\mathbf{Q}}}
 e^{i\mathbf{Q}\cdot(\mathbf{L}+\boldsymbol{\kappa}-\mathbf{L}'-\boldsymbol{\kappa}')}\nonumber\\\times&
 \boldsymbol{\epsilon}_{\alpha\boldsymbol{\kappa}\Lambda}({\mathbf
 Q})\boldsymbol{\epsilon}_{\beta\boldsymbol{\kappa}'\Lambda}({\mathbf Q})
  \mathcal{D}_{\Lambda}^{p}(\mathbf{Q};\underbar{t}_{1},\underbar{t}_{1'}).
 \end{align}
 In the case of coupling to equilibrium phonons, the scalar propagator can be expressed in a form similar to Eq.
 \eqref{eq:equilphoten}, with the occupation $N^{p}_{\Lambda,\mathbf{Q}}$ given by the Bose-Einstein distribution at the
 temperature of the heat bath.
 
 In the case of broken translational invariance caused by the use of nanostructures, the size of the
 eigenvalue problem in \eqref{eq:phonon_evp} will increase by the dimension of the unit cell in the direction of
 aperiodicity, while the dimension of the reciprocal space is correspondingly reduced. However, the system is not elastically open in the sense that a
 scattering problem needs to be solved, i.e., the vibrational state can be described in terms of vibrational eigenmodes
 in all dimensions. 

\subsubsection{Contacts}

In theory, contacts represent the open boundary conditions that need to be imposed on an open system in contact with
the environment in order to enable charge exchange between the device and the reservoir formed by the
contact. In macroscopic theories, boundary conditions are applied to the carrier density: surface charge is either zero
in the case of ohmic contacts (local charge neutrality) or fixed by the Schottky barrier height in the case of
metal-semiconductor contacts. In microscopic transport  theories, the description of contacts to open systems is itself
a whole field of ongoing research. At the most basic level, the formalism is required to
provide information about the density of states in the contact, the occupation of these states, and the coupling of the
contact 	states to the states in the device. In the NEGF formalism, all this information is contained in the boundary or contact self-energy. 

In the electronic case, the general treatment (see e.g. \cite{datta:95}) starts from the
partitioning of the overall GFs into contact and device contributions, according to the overall
Hamiltonian in \eqref{eq:hampart_cont} (in a discrete basis)
\begin{align}
  \mathbf{G}^{R}=&\begin{pmatrix}
(E+i\eta)\mathbbm{1}-{\bf H}_{D}&-\boldsymbol{\tau}\\
-\mathbf{\tau}^{\dagger}&(E+i\eta)\mathbbm{1}-\mathbf{H}_{R}
\end{pmatrix}^{-1}\\\equiv&\begin{pmatrix}
    \mathbf{G}_{D}^{R}&\mathbf{G}_{DR}^{R}\\
    \mathbf{G}_{RD}^{R}&\mathbf{G}_{R}^{R}
\end{pmatrix},
\label{eq:overallgf}
\end{align}
where $\boldsymbol{\tau}\equiv \mathbf{H}_{DR}$. Eq. \eqref{eq:overallgf} is identical to the Dyson
equation produced by a first order perturbation in the contact-device Hamiltonian, i.e. using 
$\mathcal{\hat{H}}'=\mathcal{\hat{H}}_{DR}$ in \eqref{eq:pertexp_el}. If written out in components, it can be used to
obtain the retarded device GF $\mathbf{G}_{D}^{R}$ in terms of the retarded GF
\begin{equation}\mathbf{g}^{R}\equiv\left[(E+i\eta)\mathbbm{1}-\mathbf{H}_{R}\right]^{-1}
\label{eq:leadgf}
\end{equation}
of the isolated reservoir: replacing $\mathbf{G}_{DR}^{R}=\mathbf{G}^{R}_{D}\mathbf{\tau}\mathbf{g}^{R}$,
the retarded device GF results in
\begin{align}
\mathbf{G}_{D}^{R}=&\left[(E+i\eta)\mathbbm{1}-\mathbf{H}_{D}-\boldsymbol{\tau}^{\dagger}
\mathbf{g}^{R}\boldsymbol{\tau}\right]^{-1}
\end{align}
which defines the retarded boundary self-energy 
\begin{align}
\mathbf{\Sigma}^{RB}=\boldsymbol{\tau}^{\dagger}
\mathbf{g}^{R}\boldsymbol{\tau}.
\end{align}
With the assumption of vanishing scattering between contacts and device, the Keldysh relation for
the correlation functions of device and reservoirs, in analogy to \eqref{eq:overallgf} provides the in- and outscattering contact
 self-energy terms
\begin{align}
 \mathbf{\Sigma}^{\lessgtr
B}=\boldsymbol{\tau}\mathbf{g}^{\lessgtr}\boldsymbol{\tau}^{\dagger},
 \end{align}
where $\mathbf{g}^{\lessgtr}$ is  the lesser/greater GF of the uncoupled reservoir. 
A frequently used assumption is that of contacts which are equilibrated due to scattering, which is
usually a good approximation for solar cell devices operating at elevated temperature and possessing extended contacts.
In this case, the lead correlation functions can be expressed in terms of the retarded and advanced functions using the
fluctuation-dissipation theorem,
\begin{align}
\mathbf{g}^{<}=&if_{\mu}(E)\mathbf{a},\quad\mathbf{g}^{>}=i[1-f_{\mu}(E)]\mathbf{a},
\end{align}
where $\mu$ is the chemical potential and $\mathbf{a}=i[\mathbf{g}^{R}-\mathbf{g}^{A}]$ the spectral function of the
reservoir.  

The whole problem of the coupling to the reservoirs is now reduced to the calculation of the contact
GF $\mathbf{g}^{R}$, which in principle is of infinite dimension,
but only needs to be known in the close vicinity of the device boundary, owing to the
reduced dimensionality of the coupling matrix
$\mathbf{\tau}$, and can therefore be calculated by surface GF
methods using decimation techniques \cite{guinea:83,lopez_sancho:84},
conformal maps \cite{umerski:97} or complex band methods \cite{lee:81_2,chang:82,bouwen:95,ogawa:99}. 

Similar surface GF approaches can also be used for photons \cite{jahnke:95,rahachou:05} or phonons
\cite{velasco:88,garcia_moliner:86} in systems that are open in the optical 
or vibrational sense, respectively.

\subsubsection{Interactions}%do also consider the corresponding problem, i.e. cite the literature for the photon or
%phonon problem
The effects of interactions between the different degrees of freedom can be considered
perturbatively via respective self-energy terms in the same way as the perturbation due to contacts by replacing the
perturbation Hamiltonian in \eqref{eq:pertexp_el} and \eqref{eq:pertexp_bos} with the corresponding
interaction term. Alternatively, the self-energies may be derived using  variational-derivative techniques
\cite{kadanoff:62,schaefer:02}.

\paragraph{electron-photon}%cite works by Henneberger, Haug, Pereira, Jahnke,Koch

The single particle potential for linear coupling of the electromagnetic field to the
electronic system is obtained in Coulomb-gauge as
\begin{align}
 \hat{H}_{e\gamma}(\underbar{1})=-\frac{e}{m_{0}}\hat{{\mathbf A}}(\underbar{1})\cdot\hat{{\mathbf
 p}}(1)\label{eq:elphotham}
\end{align}
with  $\hat{{\mathbf A}}$ the vector potential operator of the electromagnetic field and
$\hat{{\mathbf p}}(1)\equiv-i\hbar\nabla_{\mathbf{r}_{1}}$ the momentum operator. To first order in
the vector potential, the interaction results in the singular self-energy term
\begin{align} 
\Sigma_{e\gamma}^{\delta}(\underline{1})=-\frac{i\hbar e}{m_{0}}\langle
\hat{\mathbf{A}}(\underline{1})\rangle\cdot \hat{\mathbf{p}}(1)\equiv-\frac{i\hbar
e}{m_{0}}\hat{\mathbf{A}}_{coh}(\underline{1})\cdot \hat{\mathbf{p}}(1),\label{eq:singular_se}
\end{align}
describing the coupling to classical light, which corresponds to a \emph{mean-field} treatment of
the interaction. Since this self-energy is proportional to the coherent polarization, it leads to
off-diagonal entries if a band basis is used \cite{pereira:98}, which can be eliminated using a
band-decoupling scheme \cite{henneberger:88_3}, resulting in an effective interband self-energy that
is second-order in the field \cite{ae:negfphot_10}, e.g. for a two-band model $(b=c,v)$
\begin{align}
\tilde{\Sigma}_{e\gamma,cc}(\underline{1},\underline{1'})=&\Sigma_{e\gamma,cv}^{\delta}(\underline{1})
G_{vv}(\underline{1},\underline{1}')\Sigma_{e\gamma,vc}^{\delta}(\underline{1}').\label{eq:coh_elphotse}
\end{align}
To second order in the vector potential, the self-energy from the Fock term
of the self-consistent Born approximation is
\begin{align}
\Sigma_{e\gamma}^{F}(\underline{1},\underline{2})=&\frac{i\hbar\mu_{0}
e^{2}}{2m_{0}^{2}}\sum_{\alpha,\beta}\left[\hat{p}_{\alpha}(1)-\hat{p}_{\alpha}(1')
\right]G(\underline{1},\underline{2})\nonumber\\ &\times\hat{p}_{\beta}(2)
\overleftrightarrow{\mathcal{D}}_{\alpha\beta}(\underline{2},\underline{1'})|_{1'=1},\label{eq:fock_se}
\end{align}
where the photon GF is defined in \eqref{eq:photgf}. Close inspection of the two
expressions shows that the self-energy component in \eqref{eq:fock_se} due to the subtraction of
the coherent part is formally identical to \eqref{eq:coh_elphotse}.

In analogy to the formal treatment of the electronic system, a perturbative or variational
approach can be used to derive the photon self-energy component due to electron-photon interaction,
according to the expansion in \eqref{eq:pertexp_bos}. To lowest order, corresponding to the
\emph{random-phase-approximation} (RPA) for the polarization function, the transverse photon self-energy
reads
\begin{align}
\overleftrightarrow{\Pi}^{\gamma}_{\alpha\beta}(\underbar{1},\underbar{2})=&-\frac{i\hbar\mu_{0}e^{2}}{4
m_{0}^2}\left[\hat{p}_{\alpha}(1)-\hat{p}_{\alpha}(1')\right]G(\underbar{1},\underbar{2})\nonumber\\
&\times\hat{p}_{\beta}(2) G(\underbar{2},\underbar{1}')|_{1'=1}.
\end{align}

\paragraph{Electron-Phonon}% cite work by Fredericksen, Hylgaard, diCarlo, Luisier, Lu

The effects of the vibrational modes on the electronic degrees of freedom of the system  are
described in terms of the coupling of the force field of the electron-ion potential $V_{ei}$ to the 
quantized field $\boldsymbol{\mathcal{\hat{U}}}$ of the ionic displacement \cite{mahan:90}, 
\begin{align}
\hat{H}_{ep}({\mathbf
r},t)=&\sum_{\mathbf{L},\boldsymbol{\kappa}}\boldsymbol{\mathcal{\hat{U}}}(\mathbf{L}+\boldsymbol{\kappa},t)\cdot
\nabla V_{ei}[\mathbf{r}-(\mathbf{L}+\boldsymbol{\kappa})],\label{eq:el_phonint_rs}
\end{align}
with the displacement field given by the normal-mode expansion \eqref{eq:normalmode}.
The potential felt by electrons in heterostructure states due to coupling to \emph{bulk} phonons can thus be
written as
\begin{equation}
  \hat{H}_{ep}({\mathbf r},t)=\sum_{\Lambda{\mathbf
  Q}}\frac{U_{\Lambda,{\mathbf Q}}}{\sqrt{V}}e^{i{\mathbf Q}\cdot{\mathbf
  r}}\{\hat{b}_{\Lambda,{\mathbf Q}}(t)+ \hat{b}_{\Lambda,-{\mathbf
  Q}}^{\dagger}(t)\},
\label{eq:el_phonint}
\end{equation}
where $\mathbf{r}$ is the electron coordinate, and
$U_{\Lambda,\mathbf{Q}}$ are related to the Fourier coefficients of
the electron-ion potential $V_{ei}$ \cite{mahan:90}.
The electron-phonon self-energy can be obtained from the interaction potential
\eqref{eq:el_phonint_rs} within many-body perturbation theory in complete analogy to the
electron-photon interaction. The self-energy corresponding to the Fock term of the SCBA appears in
the form
\begin{align}
\Sigma_{ep}^{F}(\underbar{1},\underbar{2})=&\sum_{\mathbf{L},\boldsymbol{\kappa}}\sum_{\mathbf{L}',\boldsymbol{\kappa}'}
\sum_{\alpha,\beta}\mathcal{F}_{\alpha}\left(\mathbf{r}_{1}-\mathbf{L}\boldsymbol{\kappa}\right)
\mathcal{F}_{\beta}\left(\mathbf{r}_{2}-\mathbf{L}'\boldsymbol{\kappa}'\right)
\nonumber\\&\times\mathcal{D}_{\alpha\beta}^{p}(\mathbf{L}\boldsymbol{\kappa},t_{1};
\mathbf{L}'\boldsymbol{\kappa}',t_{2})G(\underbar{1},\underbar{2}),
\end{align}
with $\boldsymbol{\mathcal{F}}=\nabla V_{ei}$. For inhomogeneous systems, one has to consider also the Hartree term
\cite{hyldgaard:94}, which reads
\begin{align}
\Sigma_{ep}^{H}(\underbar{1},\underbar{2})=&\Big[\sum_{\mathbf{L},\boldsymbol{\kappa}}
\sum_{\mathbf{L}',\boldsymbol{\kappa}'} \sum_{\alpha,\beta}\int
d\underbar{3}\mathcal{F}_{\alpha} \left(\mathbf{r}_{1}-\mathbf{L}\boldsymbol{\kappa}\right) 
\mathcal{F}_{\beta}\left(\mathbf{r}_{3} -\mathbf{L}'\boldsymbol{\kappa}'\right) \nonumber\\
&\times\mathcal{D}_{\alpha\beta}^{p}(\mathbf{L} \boldsymbol{\kappa},t_{1};
\mathbf{L}'\boldsymbol{\kappa}',t_{3})G(\underbar{3},\underbar{3})\Big]\delta(\underbar{1},
\underbar{2}),
\end{align}
In the case where the phonon GFs are not computed self-consistently, it is often more
convenient to use in the derivation of the self-energy the form \eqref{eq:el_phonint} for
the interaction potential. Finally, the RPA-expression of the phonon self-energy required for the renormalization of the
phonon propagator reads
\begin{align}
\Pi_{\mu\nu}^{p}(\tilde{\underbar{1}},\tilde{\underbar{2}})=&i\hbar \int d^{3}r_{1} \int d^{3}r_{2}\mathcal{F}_{\mu}
\left(\mathbf{r}_{1}-\mathbf{L}_{1}\boldsymbol{\kappa}_{1}\right)\nonumber\\&\times\mathcal{F}_{\nu}\left(\mathbf{r}_{3}
-\mathbf{L}_{2}\boldsymbol{\kappa}_{2}\right)G(\underbar{1},\underbar{2})G(\underbar{2},\underbar{1}).
\end{align}

\paragraph{Carrier-carrier}% cite work by Henneberger, Haug, Pereira,Koch, also GW developments at
%the molecular level 
In general, there are two types of electronic interactions to be considered in solar cells:
electron-electron and electron-hole interactions. While the former affect primarily the
intraband-dynamics and are thus of relevance in the discussion of thermalization and relaxation
issues in transport, the latter can have a strong impact on the optical properties via the
excitonic enhancement of interband transitions. Since we do not consider the case of excitonic solar
cells where exciton diffusion is a key process, like in organic photovoltaics, this absorption
enhancement will be the only excitonic effect to be discussed. 

The most important effect of electronic interactions in junction-based inorganic solar cells is the
built-in potential due to bipolar doping and the modification of the electrostatic potential $U$
under illumination. Both cases can be treated macroscopically via the solution of Poisson's equation 
\begin{align}
\epsilon_{0}\nabla_{\mathbf{r}}\left[\epsilon(\mathbf{r})\nabla_{\mathbf{r}}U(\mathbf{r})\right]=n(\mathbf{r})-N_{d}(\mathbf{r}),
\label{eq:poisseq}
\end{align}
for a given profile $N_{d}$ of fully ionized dopands and (static) longitudinal dielectric function $\epsilon$ and with
the charge carrier density $n$ (including holes) from the GFs. In the microscopic
picture, the result of this approach corresponds to the Hartree-term, i.e. a mean-field level
treatment, of carrier-carrier interaction. The corresponding effective single-particle interaction potential is
\begin{align}
\hat{V}_{ee}(\mathbf{r},t)=\int
d^{3}r'\hat{\Psi}^{\dagger}(\mathbf{r}',t)V_{ee}(\mathbf{r'}-\mathbf{r})\hat{\Psi}(\mathbf{r}',t),
\end{align}
where $V_{ee}(\mathbf{r'}-\mathbf{r})$ is the two-body interaction potential. The Hartree
self-energy for this potential is 
\begin{align}
\Sigma^{H}_{ee}(\underbar{1},\underbar{2})=&i\hbar\delta(\underbar{1},\underbar{2})\int_{\mathcal{C}}
d\underbar{2}V_{ee}(\mathbf{r}_{1}-\mathbf{r}_{2})G(\underbar{2},\underbar{2})\\
\equiv& \delta(\underbar{1},\underbar{2})U(\underbar{1}),
\end{align}
where $U$ is the mean-field potential. On the same level of perturbation theory, the corresponding
Fock term reads
\begin{align}
\Sigma^{F}_{ee}(\underbar{1},\underbar{2})=&i\hbar\delta(\underbar{t}_{1}-\underbar{t}_{2})V_{ee}(\mathbf{r}_{1}-\mathbf{r}_{2})
G(\underbar{1},\underbar{2}).
\end{align}

Electron-electron interactions beyond the mean-field Hartree-Fock level may be considered by
explicit inclusion of dynamical screening via the $GW$ formalism. For practical purposes, often a
statically screened potential is assumed together with the above forms of the self-energy. 
Concerning the electron-hole interaction beyond Poisson's equation, excitonic effects in
semiconductor nanostructures have been discussed for steady state linear absorption or in the
regimes of high pulse excitations. For the latter, sophisticated quantum-kinetic theories were developed
\cite{haug:88,haug:96,haug:04,henneberger:88_3,henneberger:96,jahnke:95}.
For extended systems, the computational cost of such approaches limits their applicability to
special cases of reduced complexity, such as systems that may be described with a few eigenstates, 
and they are thus not suited for the investigation of transport properties. For this reason,
excitonic	 effects will be considered here only in terms of the coulomb-enhancement of the optical 
interband self-energy due to the \emph{coherent} polarization of the semiconductor system resulting 
from the coupling to a coherent (classical) light field, as introduced in \eqref{eq:coh_elphotse}. 
The singular interband Coulomb term is ($a\neq b=c,v$)
\begin{align}
\Sigma_{ee,ab}^{\delta}(\mathbf{r}_{1},\mathbf{r}_{2},t)=i\hbar
V(\mathbf{r}_{1}-\mathbf{r}_{2})G_{ab}^{<}(\mathbf{r}_{1},\mathbf{r}_{2},t,t^{+}),\label{eq:cb_sing_se}
\end{align}
where $V$ is the (screened) Coulomb potential, and depends thus on the coherent
interband polarization through the interband GF, which the decoupling provides in the form
\begin{align}
G_{vc}^{<}(\underline{1},\underline{1'})&=\int_{\mathcal{C}} d\underbar{2}\int_{\mathcal{C}}
d\underbar{3}\,\Big[\tilde{G}_{vv}^{R}(\underline{1},\underline{2})\Sigma_{vc}^{\delta}(\underline{2},\underline{3})
G_{cc}^{<}(\underline{3},\underline{1'})\nonumber\\&+\tilde{G}_{vv}^{<}(\underline{1},\underline{2})
\Sigma_{vc}^{\delta}(\underline{2},\underline{3}) G_{cc}^{A}(\underline{3},\underline{1'})\Big]\\ 
\equiv&-\frac{i}{\hbar}\int_{\mathcal{C}} d 2\int_{\mathcal{C}}
d3\,\Sigma_{vc}^{\delta}(\underline{2},\underline{3})\mathcal{T}_{vc}(\underline{1},\underline{2},
\underline{3},\underline{1}'),
\end{align}
with the tilde denoting the GF without interband coupling. Inserting the explicit expressions for the singular
self-energies leads to the Bethe-Salpeter equation (BSE) for the coherent polarization, which for steady state
reads
\begin{align}
G_{vc}^{<}(\mathbf{r}_{1},\mathbf{r}_{1'};E)&=G_{vc,(0)}^{<}(\mathbf{r}_{1},\mathbf{r}_{1'};E)\nonumber\\
+&\int d^{3}r_{2}\int
d^{3}r_{3}V(\mathbf{r}_{2}-\mathbf{r}_{3})\nonumber\\
\times&\mathcal{T}_{vc}(\mathbf{r}_{1},\mathbf{r}_{2},
\mathbf{r}_{3},\mathbf{r}_{1'};E) G_{vc}^{<}(\mathbf{r}_{2},\mathbf{r}_{3},E),\label{eq:BSE_realspace}
\end{align}
with
\begin{align}
G_{vc,(0)}^{<}(\mathbf{r}_{1},\mathbf{r}_{1'};E)=&-\frac{ie}{m_{0}\hbar}\int
d^{3}r_{2}\hat{\mathbf{A}}_{coh}(\mathbf{r}_{2},E)\cdot
\hat{\mathbf{p}}(\mathbf{r}_{2})\nonumber\\&\times\mathcal{T}_{vc}(\mathbf{r}_{1},
\mathbf{r}_{2},\mathbf{r}_{2},\mathbf{r}_{1'};E)\label{eq:polfun_coh}
\end{align}
the coherent interband polarization of non-interacting electron-hole pairs.
The corresponding self-consistent BSE-type equation for the singular self-energy is obtained by
using \eqref{eq:BSE_realspace} and \eqref{eq:polfun_coh} in Eq. \eqref{eq:cb_sing_se},
\begin{align}
&\Sigma_{ab}^{\delta}(\mathbf{r}_{1},\mathbf{r}_{1}',E)=\Sigma_{ab,(0)}^{\delta}(\mathbf{r}_{1},\mathbf{r}_{1}',E)
+V(\mathbf{r}_{1}-\mathbf{r}_{1}')\int d^{3}r_{2}\nonumber\\&\times\int
d^{3}r_{3}\mathcal{T}_{vc}(\mathbf{r}_{1},
\mathbf{r}_{2},\mathbf{r}_{3},\mathbf{r}_{1'};E)\Sigma_{ab}^{\delta}(\mathbf{r}_{2},\mathbf{r}_{3},E),
\end{align}
where $\Sigma_{ab,(0)}^{\delta}=\Sigma_{ab}^{e\gamma,\delta}$.

\subsection{Steady state device characteristics from NEGF}
Due to the very general grounds of their determination via quantum kinetic equations, the NEGF contain information not
only about the electronic, optical and vibrational states of the system, but also about the occupation of these states
and the rates of transitions between them. Knowledge of the GFs for carriers, photons and phonons thus 
immediately provides information about the device characteristics as they are related directly to physical observables: 
the ensemble average of any single-body operator  $\hat{\mathcal{O}}$ can be written as \cite{fetter_walecka}
\begin{align}
\langle \hat{\mathcal{O}}(\mathbf{r},t)\rangle&=\mathrm{Tr}[\rho
\hat{\mathcal{O}}(\mathbf{r},t)]\nonumber\\ 
&=\mp \lim_{\mathbf{r}\rightarrow\mathbf{r}}\lim_{t'\rightarrow t+}\mathrm{Tr}[O(\mathbf{r},t)
\mathcal{F}^{<}({\mathbf r},t;{\mathbf r'},t')],
\label{eq:relgfopaverage}
\end{align}
where $\mathcal{F}=G,\mathcal{D}^{\gamma,p}$ and the upper (lower) sign applies to fermions (bosons).

In photovoltaics, the properties of interest determined experimentally cover on the one hand a range of \emph{spectral}
quantities, such as absorption, emission and transmission spectra, or spectral response and quantum
efficiency. These quantities can be obtained from the GF in the energy (frequency) domain. On the
other hand, there are the \emph{integral} quantities which follow from the corresponding spectral versions upon energy
integration, such as the charge carrier current density providing the photovoltaic current-voltage $(J/V)$
characteristics which determine the energy-conversion efficiency of the solar cell
device.

\subsubsection{Density and current}
Following the above procedure, the electron density in steady-state is computed by inserting the electron density
operator $\hat{\rho}({\mathbf r},t)=\hat{\Psi}^{\dagger}({\mathbf r},t)\hat{\Psi}({\mathbf r},t)$ in Eq.
\eqref{eq:relgfopaverage} and performing the Fourier-transform to the energy domain,
\begin{align}
n_{e}({\mathbf r})=&- i\int \frac{dE}{2\pi}G^{<}({\mathbf r},{\mathbf r};E).
 \label{eq:steaddens}
  \end{align}
The hole density is obtained by replacing in the above expression $G^{<}$ with $-G^{>}$.
In the same way, the photon correlation function yields the photon density via
\begin{align}
n_{\gamma}(\mathbf{r})=i\int\frac{dE}{2\pi}\mathrm{Tr}\{\boldsymbol{\mathcal{D}}^{\gamma
<}(\mathbf{r},\mathbf{r},E)\},
\end{align}
where the integrand is given by Planck's law if the correlation function for  equilibrium free-field
photons is used.

The electronic current density is determined as the ensemble average of the operator
\begin{align}
\hat{\mathbf{j}}_{e}(\mathbf{r},t)=&\frac{-ie\hbar}{2m_{0}}\left[\hat{\Psi}^{\dagger}(\mathbf{r})
\nabla_{\mathbf{r}}\hat{\Psi}(\mathbf{r})-\{\nabla_{\mathbf{r}}\hat{\Psi}^{\dagger}(\mathbf{r})\}
\hat{\Psi}(\mathbf{r})\right]\\
&\equiv\lim_{\mathbf{r'}\rightarrow\mathbf{r}}\frac{-ie\hbar}{2m_{0}}\left[\nabla_{\mathbf{r}}
-\nabla_{\mathbf{r'}}\right]\hat{\Psi}^{\dagger}(\mathbf{r})\hat{\Psi}(\mathbf{r}'),
\label{eq:negfcurr}
\end{align}
and in steady state can be expressed in terms of the carrier GF via
\begin{equation}
\mathbf{j}_{e}(\mathbf{r})=
\lim_{\mathbf{r'}\rightarrow\mathbf{r}}\frac{\hbar}{2m_{0}}[\nabla_{\mathbf{r}}
-\nabla_{\mathbf{r'}}]\int \frac{dE}{2\pi}G^{<}(\mathbf{r},\mathbf{r'};E).
\end{equation}
This form of the current may also be obtained directly from the formulation of the electron
continuity equation in terms of the electronic GFs by inspection of the divergence
term.

The photon current $\mathbf{j}_{\gamma}$ in the
device follows from the evolution of the light intensity, which is given by the symmetrized non-equilibrium ensemble
average of the Poynting vector operator \cite{richter:08},
\begin{align}
|\langle\mathbf{S}(\mathbf{r}_{1})\rangle|=&\int_{0}^{\infty}
d(\hbar\omega)\hbar\omega ~\mathbf{j}_{\gamma}(\mathbf{r}_{1},\hbar\omega),
\end{align}
where
\begin{align}
\langle\mathbf{S}(\mathbf{r}_{1},t_{1})\rangle \equiv&\frac{1}{\mu_{0}}\Big\langle
\hat{\boldsymbol{\mathcal{E}}}(\mathbf{r}_{1},t_{1})
\times\hat{\boldsymbol{\mathcal{B}}}(\mathbf{r}_{2},t_{2})\Big\rangle\Big|_{sym,1=2},
\end{align}
with
\begin{align}
\hat{\boldsymbol{\mathcal{E}}}(\mathbf{r},t)=-\frac{\partial
\hat{\mathbf{A}}(\mathbf{r},t)}{\partial t},\quad
\hat{\boldsymbol{\mathcal{B}}}(\mathbf{r},t)=\nabla\times\hat{\mathbf{A}}(\mathbf{r},t)
\end{align}
the electric and magnetic components of the EM field. Recalling the definition \eqref{eq:photgf} of
the photon GF, the spectral photon current results in terms of the latter to
\begin{align}
j_{\gamma,\mu}(\mathbf{r}_{1},\hbar\omega)=&\lim_{\mathbf{r}_{2}\rightarrow\mathbf{r}_{1}}
\frac{1}{\mu_{0}}\sum_{\nu\neq \mu}\mathrm{Re}\Big[\nabla_{\nu}(2)
\big\{\mathcal{D}_{\nu\mu}^{>}(\mathbf{r}_{1},\mathbf{r}_{2};\hbar\omega)\nonumber\\
&+\mathcal{D}_{\nu\mu}^{<}(\mathbf{r}_{1},\mathbf{r}_{2};\hbar\omega)\big\}
-\nabla_{\mu}(2)\big\{\mathcal{D}_{\nu\nu}^{>}
(\mathbf{r}_{1},\mathbf{r}_{2};\hbar\omega)\nonumber\\
&+\mathcal{D}_{\nu\nu}^{<}(\mathbf{r}_{1},
\mathbf{r}_{2};\hbar\omega)\big\}\Big].
\end{align}

\subsubsection{Conservation laws and scattering rates}
 The macroscopic balance equation for a photovoltaic system is
the steady state continuity equation for the charge carrier density
\begin{align} 
\nabla\cdot 
{\mathbf j}_{c}({\mathbf r})=\mathcal{G}_{c}(\mathbf{r})-\mathcal{R}_{c}(\mathbf{r}),\quad
c=e,h, 
\label{eq:macro_cont}
\end{align}
where $j_{c}$ is the particle current density, $\mathcal{G}_{c}$ the (volume) generation rate and
$\mathcal{R}_{c}$ the recombination rate of carriers species $c$ .
In the microscopic theory, the divergence of the (particle) current density is given by 
\cite{kadanoff:62,keldysh:65}
\begin{align}
\nabla\cdot 
\mathbf{j}(\mathbf{r})
&=-\frac{2}{V}\int\frac{dE}{2\pi\hbar}\int d^3 r'\Big[\Sigma^{R}
(\mathbf{r},\mathbf{r}';E)G^{<}(\mathbf{r}',\mathbf{r};E)
\nonumber\\+&\Sigma^{<}(\mathbf{r},\mathbf{r}';E)G^{A}(\mathbf{r}',\mathbf{r};E)
-G^{R}(\mathbf{r},\mathbf{r}';E)\nonumber\\\times&\Sigma^{<}(\mathbf{r}',\mathbf{r};E)-
G^{<}(\mathbf{r},\mathbf{r}';E)\Sigma^{A}(\mathbf{r}',\mathbf{r};E)\Big].\label{eq:currcons}
\end{align}
The RHS of the above equation provides a general expression for the microscopic intra- and
interband rates of (potentially non-local) scattering processes. Current conservation requires the
RHS to vanish for energy integration over all bands or states, which means that any scattering 
contributions originating in transitions in energy space must cancel upon energy integration. If
the integration is restricted to either conduction or valence bands, the above equation corresponds
to 	the microscopic version of \eqref{eq:macro_cont} and provides on the RHS the total local
interband scattering rate. The total interband current is found by integrating the carrier	
specific divergence over the active volume, and is equivalent to the total global transition rate
and, via the Gauss theorem, to the difference of the interband currents at the boundaries of the
interacting region. Making use of the cyclic property of the trace, it can be expressed in the form
\begin{align}
\mathcal{R}_{tot}=&\frac{2}{V}\int d^3 r\int\frac{dE}{2\pi\hbar}\int d^3
r'\Big[\Sigma^{<}(\mathbf{r},\mathbf{r}';E)G^{>}(\mathbf{r}',\mathbf{r};E)\nonumber\\
&-\Sigma^{>}(\mathbf{r},\mathbf{r}';E)G^{<}(\mathbf{r}',\mathbf{r};E)
\Big]\label{eq:totrate}
\end{align}
with units $[\mathcal{R}]=s^{-1}$. The above equation can be used to define the spectral scattering
current $J_{\phi}(E)$ via $\mathcal{R}_{\phi}=\int dE J_{\phi}(E)$ for a given scattering process
described by a self-energy $\Sigma_{\phi}$, such that \cite{datta:95}
\begin{align}
J_{\phi}(E)=&\frac{1}{2\pi\hbar}\int d^3 r\int d^3
r'\Big[\Sigma^{<}_{\phi}(\mathbf{r},\mathbf{r}';E)G^{>}(\mathbf{r}',\mathbf{r};E)\nonumber\\
&-\Sigma^{>}_{\phi}(\mathbf{r},\mathbf{r}';E)G^{<}(\mathbf{r}',\mathbf{r};E) \Big],\\
\equiv&\frac{1}{2\pi\hbar}\mathrm{tr}\Big[\Sigma^{<}_{\phi}(E)G^{>}(E)-\Sigma^{>}_{\phi}(E)G^{<}(E)
\Big],
\end{align} 
which describes the total scattering rate at a given energy as difference of the inscattering rate
into an empty state at a that energy
$\hbar^{-1}\mathrm{tr}\big[\Sigma^{<}_{\phi}(E)G^{>}(E)\big]$ and the outscattering rate 
$\hbar^{-1}\mathrm{tr}\big[\Sigma^{<}_{\phi}(E)G^{>}(E)\big]$ from an occupied state at the same 
energy. 

If we are interested in the interband scattering rate, we can neglect in Eq.
\eqref{eq:totrate} the contributions to the self-energy from intraband scattering current, e.g. via
interaction with phonons, low energy photons (free carrier absorption) or ionized impurities, since 
they cancel upon energy integration over the band. However, the current flow in a given band is not 
unaffected by the intraband scattering, since the carrier GFs are modified by the
action of the corresponding self-energies in the associated Dyson equations. 

Due to its generality, the above representation of the global current components in relation to the
carrier self-energies has manifold applications, depending on the nature of the self-energies. First
of all, it can be used to express the terminal current at contact $\alpha$ in terms of the contact
self-energies, 
\begin{align}
J_{\alpha}=&\int\frac{dE}{2\pi\hbar}\mathrm{tr}\Big[\Sigma^{B<}_{\alpha}(E)G^{>}(E)-\Sigma^{B>}_{\alpha}(E)G^{<}(E)
\Big]
\end{align}
For an equilibrium contact characterized by a chemical potential $\mu_{\alpha}$ and broadening
function $\Gamma_{\alpha}^{B}\equiv i(\Sigma^{B>}_{\alpha}-\Sigma^{B<}_{\alpha})$, the above current
formula results in the Meir-Wingreen expression
\cite{meir:92} 
\begin{align}
J_{\alpha}=&\int\frac{dE}{2\pi\hbar}\mathrm{tr}\Big[\Gamma_{\alpha}^{B}(E)\left\{f_{\mu_{\alpha}}(E)A(E)-G^{<}(E)\right\}
\Big].
\end{align}
Another important application with relevance in the simulation of solar cells, especially for the
so-called hot carrier devices, is the description of energy loss in electronic
transport via inelastic scattering processes such as coupling to phonons \cite{pecchia:07}, where 
the dissipated power is described in terms of the inelastic scattering current via \cite{datta:95}
\begin{align}
P_{e}^{diss}=-2\int dE J_{\phi}(E)E.
\end{align}
But most importantly, the formalism provides the optoelectronic conservation laws in terms of the optical rates and
associated currents via insertion of the electron-photon coupling self-energies in \eqref{eq:totrate}. On the other
hand, the global radiative interband scattering rate equals the total optical rate $R_{\gamma}^{opt}$, which, via the quantum statistical average of the
Poynting vector operator or directly from the explicit electron-photon self-energy terms in
\eqref{eq:totrate}, can be expressed in terms of photon GFs and self-energies
\cite{richter:08,henneberger:09},
\begin{align}
R_{\gamma}^{opt}&=\int d(\hbar\omega)\frac{1}{V}\int d^3 r \int d^3 r'\sum_{\mu\nu}
\Big[\overleftrightarrow{\Pi}_{\mu\nu}^{\gamma,
>}(\mathbf{r},\mathbf{r}',\hbar\omega)\nonumber\\
\times&\overleftrightarrow{D}_{\nu\mu}^{\gamma,<}(\mathbf{r'},\mathbf{r},\hbar\omega)
-\overleftrightarrow{\Pi}_{\mu\nu}^{\gamma,<}(\mathbf{r},\mathbf{r}',\hbar\omega)
\overleftrightarrow{D}_{\nu\mu}^{\gamma,>}(\mathbf{r}',\mathbf{r},\hbar\omega)\Big]\label{eq:photrate_opt}
\end{align}
and which is related to the global electromagnetic power dissipation through 
\begin{align}
P_{\gamma}^{diss}=-\int_{0}^{\infty}\frac{d\omega}{2\pi}\hbar\omega R^{opt}_{\gamma}(\hbar\omega).
\end{align}
The two terms in Eq. \eqref{eq:photrate_opt} provide microscopic expressions for absorbed
and emitted photon flux, which may be regrouped to write the optical rate in terms of the \emph{net}
absorption, i.e. the absorption minus the stimulated emission, and the \emph{spontaneous} emission,
\begin{align}
R_{\gamma,abs}^{opt}(\hbar\omega)=&\frac{1}{V}\mathrm{tr}\big[\hat{\overleftrightarrow{\Pi}}{}^{\gamma}(\hbar\omega)
\overleftrightarrow{\mathcal{D}}^{\gamma,<}(\hbar\omega)\big],\\ 
R_{\gamma,em}^{opt}(\hbar\omega)=&\frac{1}{V}\mathrm{tr}\big[\overleftrightarrow{\Pi}^{\gamma,
<}(\hbar\omega)\hat{\overleftrightarrow{\mathcal{D}}}
{}^{\gamma}(\hbar\omega)\big],
\end{align}
where the trace is over spatial coordinates and the cartesian components of the vector potential,
and we have introduced the spectral functions 
\begin{align}
\hat{\overleftrightarrow{\Pi}}{}^{\gamma}=\overleftrightarrow{\Pi}^{\gamma,>}-\overleftrightarrow{\Pi}^{\gamma,<},\quad 
\hat{\overleftrightarrow{\mathcal{D}}}{}^{\gamma}=\overleftrightarrow{\mathcal{D}}^{\gamma,>}
-\overleftrightarrow{\mathcal{D}}^{\gamma,<}
\end{align}
for electron-hole pairs and for photons, respectively.
The above expressions for $R^{opt}_{\gamma,abs/em}$ may be interpreted as global generation and
recombination rates \cite{richter:08}. The emission term provides a more general formulation of
generalized Kirchhoff law \cite{wuerfel:82} normally used in photovoltaics for the photon flux
emitted from an excited semiconductor. It allows for the inclusion of a realistic electronic dispersion
via the polarization function $\overleftrightarrow{\boldsymbol{\Pi}}^{\gamma}$, reflecting the
effects of electron-electron, electron-hole and electron-phonon interactions and the non-equilibrium
occupation of these states, as well as the consideration of the optical modes specific to the geometry of the system.

In the electronic picture, the separation of the different contributions is not
similarly straightforward, since the in- and outscattering self-energies always contain both an 
absorption and an emission component, e.g. interband absorption and intraband (free-carrier)
emission in the case of the electron inscattering self-energy. If the intraband 
terms are neglected, absorption and emission terms can be identified with the in- and outscattering
contributions in \eqref{eq:totrate}, where now the emission comprises both stimulated and
spontaneous components. 

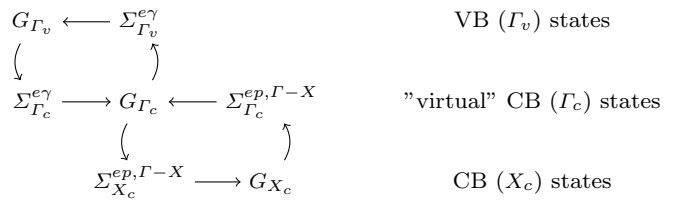
\begin{figure}[t]
 \begin{center}
 \begin{tikzpicture}
\matrix(m)[matrix of math nodes, row sep=5mm, column sep=3mm, text height=2.5mm,
text depth=1mm] {
G_{\Gamma_{v}}  \pgfmatrixnextcell \Sigma^{e\gamma}_{\Gamma_{v}}
\pgfmatrixnextcell ~ \pgfmatrixnextcell~\pgfmatrixnextcell\textnormal{VB
($\Gamma_{v}$) states} \\ \Sigma^{e\gamma}_{\Gamma_{c}}
\pgfmatrixnextcell
G_{\Gamma_{c}} \pgfmatrixnextcell \Sigma^{ep,\Gamma-X}_{\Gamma_{c}}
\pgfmatrixnextcell~\pgfmatrixnextcell\textnormal{''virtual'' CB ($\Gamma_{c}$)
states}\\ ~\pgfmatrixnextcell \Sigma^{ep,\Gamma-X}_{X_{c}}
\pgfmatrixnextcell G_{X_{c}}\pgfmatrixnextcell \pgfmatrixnextcell \textnormal{CB ($X_{c}$) states}\\};
\path[->]
(m-1-2) edge  [auto] (m-1-1)
(m-1-1) edge  [bend right] (m-2-1) 
(m-2-2) edge  [bend right] (m-1-2)
(m-2-1) edge  [auto] (m-2-2)
(m-2-3) edge  [auto] (m-2-2)
(m-2-2) edge  [bend right] (m-3-2)
(m-3-2) edge  [auto] (m-3-3)
(m-3-3) edge  [bend right] (m-2-3);
\end{tikzpicture}
\caption{Self-consistent computation of Green's functions and scattering self-energies in silicon enabling the description of phonon-assisted
indirect optical transitions \cite{ae:prb_11}. A similar procedure can be
used for the simulation of non-radiative recombination via defects by
replacing the virtual with a real midgap state.\label{fig:selfconsit}}
\end{center}
\end{figure}

In the electronic expression for the optical rate, the GFs correspond to the final state
of the scattering process, but the self-energy contains the GFs of the initial
state only in the case of direct optical transitions. In the case of indirect, phonon assisted
transitions such as those in indirect band gap semiconductors, the scattering process proceeds via
an intermediate, virtual state and the associated GFs and self-energies, as shown in Fig.
\ref{fig:selfconsit} for the example of a simple three-band model of silicon \cite{ae:prb_11}. 

Finally, the general formalism to treat scattering within the NEGF formalism can also be used to
find an approach to the description of non-radiative recombination, e.g. via defects, which
dominates in most photovoltaic devices. While an exact microscopic treatment is possible only in
special cases, a more phenomenological treatment proceeds along the lines of the general formalism
introduced above. In order to describe non-radiative recombination via defects, the scattering
induced renormalization of the GFs of carriers in localized as well as  extended (bulk) 
states needs to be quantified by the corresponding self-energy expressions, which are
\begin{align}
 \Sigma_{db}^{\lessgtr}(E)=&\sum_{\mathbf{k}}\int \frac{d\tilde{E}}{2\pi\hbar}
 \mathcal{M}_{bd}^{capt/em}(\mathbf{k},\tilde{E})G^{\lessgtr}_{b}(\mathbf{k},E\pm\tilde{E}),\nonumber\\
\Sigma_{bd}^{\lessgtr}(\mathbf{k},E)=&\int \frac{d\tilde{E}}{2\pi\hbar}
 \mathcal{M}_{cd}^{em/capt}(\mathbf{k},\tilde{E})G^{\lessgtr}_{d}(E\mp\tilde{E}), 
\end{align}
where $G_{d}$ is the GF of the defect or midgap state, $\mathcal{M}$ the matrix element of
carrier capture or emission by the defect, and $b=c,v$ for coupling of the defect to conduction and
valence band states, respectively, and the upper sign corresponds to the former case. Inserting these 
self-energy terms into the general expression for the electron-capture rate yields
\begin{align}
&R_{c\rightarrow d}=\sum_{\mathbf{k}}\int
\frac{dE}{2\pi\hbar}\Sigma^{>}_{cd}(\mathbf{k},E)G^{<}_{c}(\mathbf{k},E)\label{eq:bandstate_defrate}\nonumber\\
&=\sum_{\mathbf{k}}\int
\frac{dE}{2\pi\hbar}\int  \frac{d\tilde{E}}{2\pi\hbar}
 \mathcal{M}_{cd}^{capt}(\mathbf{k},\tilde{E})G^{>}_{d}(E-\tilde{E})G^{<}_{c}(\mathbf{k},E)\nonumber\\
&\equiv\int\frac{dE'}{2\pi\hbar}G^{>}_{d}(E')\Sigma_{cd}^{<}(E')= R_{d\rightarrow c},
\end{align}
expressing the fact that the rate of scattering out of the extended conduction band states by
carrier capture into defects equals the rate of scattering into the defect state. Using quasi-equilibrium
approximation for the carrier GFs, the defect recombination rate derived by Shockley,
Read and Hall \cite{shockley:52,hall:52} is recovered.       

\section{Numerical implementation}
For the numerical evaluation of the dynamical equations  \eqref{eq:dyson_phot}, GFs and
self-energies need to be represented in a basis of finite dimension via a suitable expansion of the
corresponding field operators, which converts the former in a linear system to be solved self-consistently with the
diverse self-energy equations  and the Poisson equation \eqref{eq:poisseq} for the electrostatic potential. Since the
specific implementation depends strongly on the device configuration under consideration, some rather general issues
arising in the numerical treatment of the NEGF formalism for solar cells will be discussed here, as well as the
implications of commonly used approximations.

\subsection{Basis}
There are several aspects that may have an impact on the specific choice of basis for the carrier field operators.
First of all, there is the electronic structure of the material system: while direct gap semiconductors like the III-V materials commonly used in
optoelectronic applications are reasonably well described by $\mathbf{k}\cdot \mathbf{p}$ approaches, the
situation is more demanding in indirect semiconductor such as silicon-based materials often used in photovoltaics,
wherefore atomistic approaches such as empirical or ab-initio tight-binding are preferred.  For molecular devices like
organic solar cells, ab-initio, quantum chemical methods such as density-functional theory are required for the
electronic-structure calculation. There is an important difference between light-emitting optoelectronic
devices and solar cells to consider at this point: while both transport and emission take place in a narrow energetic
range around the band edge, absorption of solar radiation is a broadband process. Hence, in principle, a full-zone description of the electronic
structure is required to cover the photogeneration induced by the entire solar spectrum.

Apart from the range of energy states described, the largest impact of the choice of basis concerns the spatial
resolution and flexibility therein. While effective continuum approaches such as EMA or $\mathbf{k}\cdot \mathbf{p}$
lend themselves to optimized adaptive spatial grids, allowing for coarse-grained treatment of spatially homogeneous regions such as bulk leads and
high resolution close to heterointerface, the grid is fixed by the atomic lattice in the case of an atomistic basis,
which precludes the description of extended systems, unless one resorts to multiscale approaches
\cite{a_d_maur:08}. 

In some cases, a mode space basis may be preferred over a real space basis, i.e. in the case where only a few modes
contribute to the device characteristics. This is for instance the case for the transverse modes in nanowire transport
\cite{jin:06_2}, or in periodic superstructures \cite{wacker:02}. However, in most cases, the mode spectrum in
leads or contacts deviates strongly from the discrete mode spectrum of the low dimensional absorber,
or, like e.g. in quantum well solar cells, there is a continuous transition from a discrete to a continuous mode spectrum in the energy range of
interest. Indeed, the ability to treat on equal footing bound state to bound state and bound state to continuum
transitions represents a considerable advantage of real space approaches \cite{henrickson:02}. 
  
\subsection{Energy discretization}
Among the numerical issues arising and approximations imposed due to computational limitations, the most prominent ones
for the photovoltaic applications in mind are inadequate energy mesh discretization and the non-locality of the interactions. The
former problem arises whenever resonances appear in the transmission function of a structure, either due to localization
effects leading to the formation of quasibound states or due to van Hove singularities in the density of states. 
The mesh should be refined around the resonances in order to avoid large errors in the energy integration and resulting
convergence problems. Different approaches have been developed to tackle the refinement issue, mainly adaptive
algorithms \cite{kubis:09,pourfath:09} or methods for the resolution of the poles of the GFs giving
rise to the resonances \cite{bouwen:95}. In the case where a large number of poles are present, adaptive
refinement according to the resolution of the spectral output quantities is preferrable. However, in the presence of
inelastic interactions such as coupling to broadband illumination, the situation may arise that \emph{any} two energies
are coupled, extending the need for mesh refinement to the whole energy range and thus making the use of a
coarse-grained starting mesh questionable. In any case of an irregular grid, care has to be taken to use interpolation
schemes that guarantee current conservation.

\subsection{Solution of the linear system for non-local interactions}
The determination of the full GF provides information on propagation and correlations between any
two sites of the dievice and thus includes all non-local phenomena. However, such a computation is
very time and memory consuming, wherefore approximate schemes with a limited number of non-zero off-diagonal GF elements are used, which then allows
for the application of fast computational routines such as the recursive GF algorithm \cite{lake:97}
but at the same time reducing the range of non-locality taken into account by the simulation. While the consequences are
sizable even for rather localized intraband scattering mechanisms such as coupling to phonons
\cite{kubis:06_2,steiger:thesis}, they are severe in the case of the interband scattering due to coupling to photons,
since in the latter case, due to the comparably long wave length of the light,  the non-locality is not restricted by
the matrix element of  the coupling itself, but by the density of states which couple
\cite{pourfath:09,steiger:iwce_09}. Thus, non-locality is only limited in the case of optical transitions between
localized states, as in weakly coupled quantum dot or quantum well absorbers.

\section{Applications} 
\begin{figure}[b]
\begin{center}
\includegraphics[height=6cm]{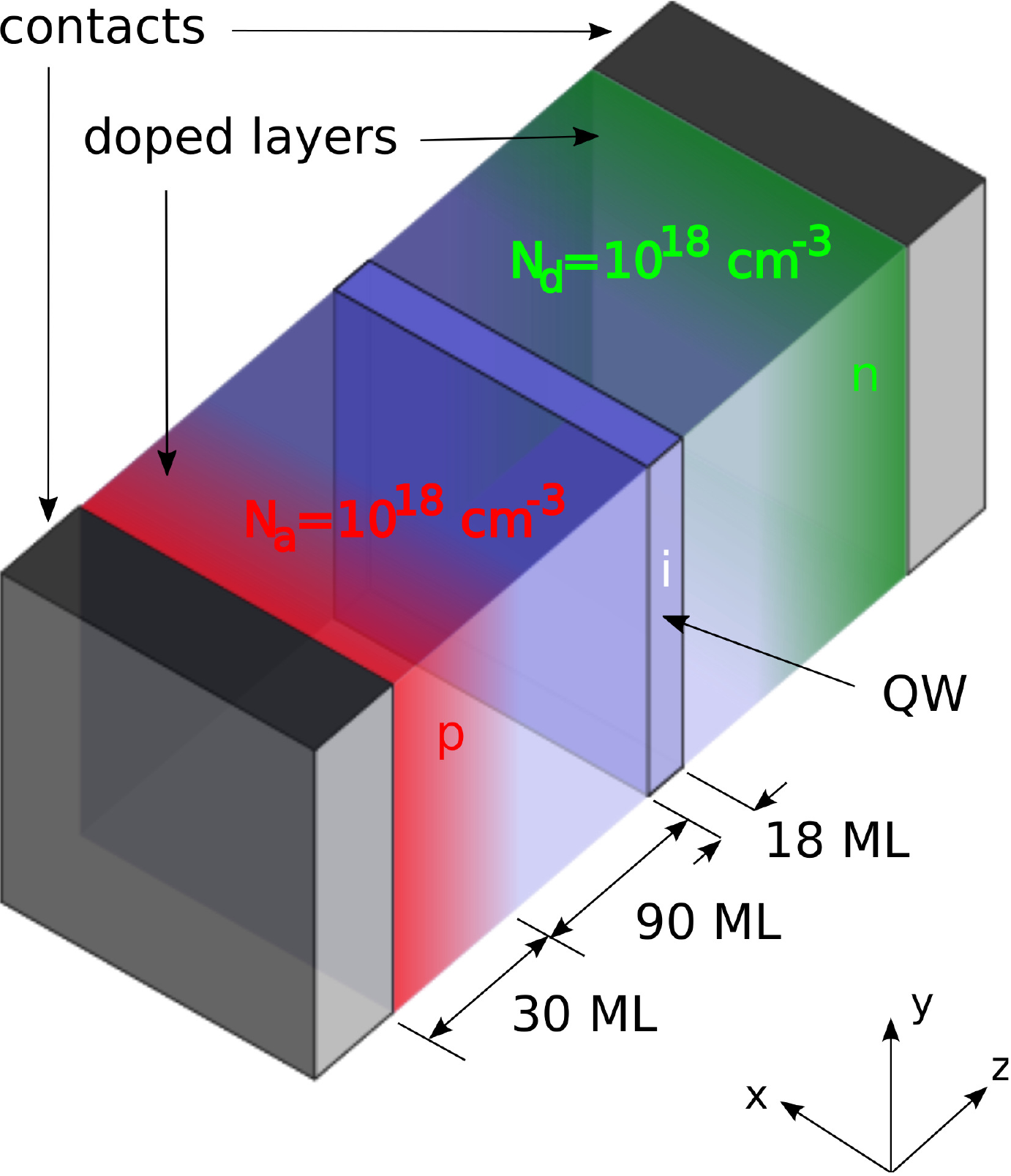}%\qquad\includegraphics[width=7cm]{figures/sqw_schematic.pdf}
\caption{Single quantum well $p$-$i$-$n$ diode configuration used in the simulation\label{fig:sqw_diode_structure}.}
\end{center}
\end{figure}
In the following, the kind of information provided by the NEGF formalism shall be illustrated by application to the simulation of two different generic quantum photovoltaic devices, namely a standard III-V material
based single quantum-well photodiode \cite{ae:prb_08,ae:spie10,ae:solmat_10} and a silicon based
superlattice absorber \cite{ae:nrl_11}. While the former device embodies all relevant structural and mechanistic
features of the multi-quantum well solar cell, the latter corresponds to a
prototypical ingredient of third-generation multi-junction devices. The two devices are
fundamentally different in several aspects of their working principles. First of all, in the
QWSC, the confined states act only as absorbers, while transport takes place in the extended
quasi-continuum states of the host material. In the superlattice device, both absorption
and transport proceed via the states of the coupled quantum wells. A second important difference is 
the role of scattering with phonons: while this interaction is a crucial aspect of the carrier
escape mechanism in the QWSC, it greatly enhances transport in the superlattice at finite internal
field due to restoration of resonant interwell charge transfer. It is thus instructive to have a
closer look at the microscopic picture of charge carrier photogeneration and transport in these two
different devices, provided by the NEGF formalism in the form of spectral currents and rates.
In both examples, coupling to equilibrium bulk phonons and isotropic free field photons is assumed.

\begin{figure}[b]
\begin{center}
\includegraphics[height=6cm]{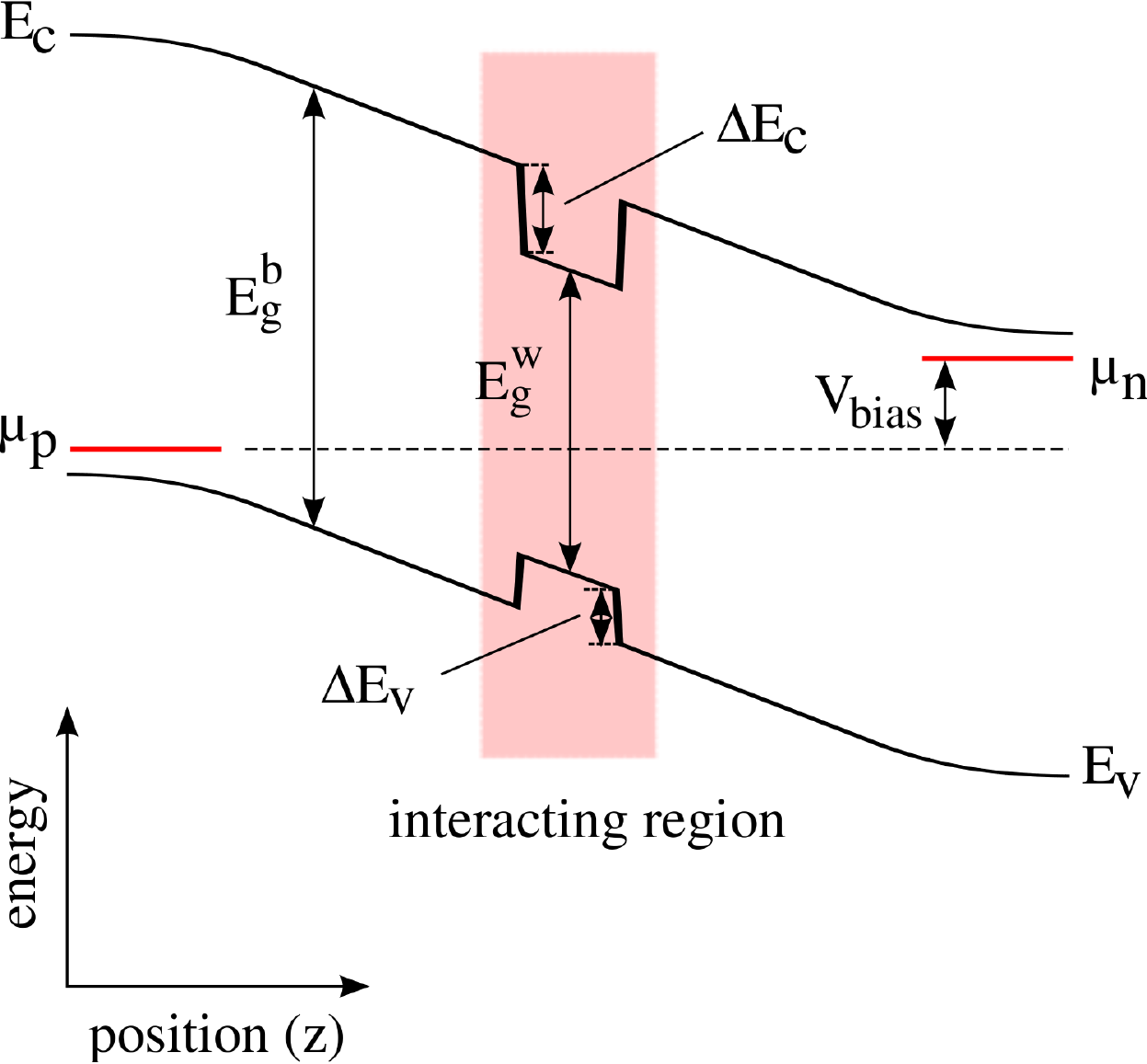}
\caption{Schematic band diagram of the device shown in Fig.\ref{fig:sqw_diode_structure}
\cite{ae:solmat_10}.\label{fig:bandscheme}}
\end{center} 
\end{figure}
\subsection{Carrier escape in direct gap single quantum well pin-diodes}  

\begin{figure*}[t]
\begin{center}
\includegraphics[width=18cm]{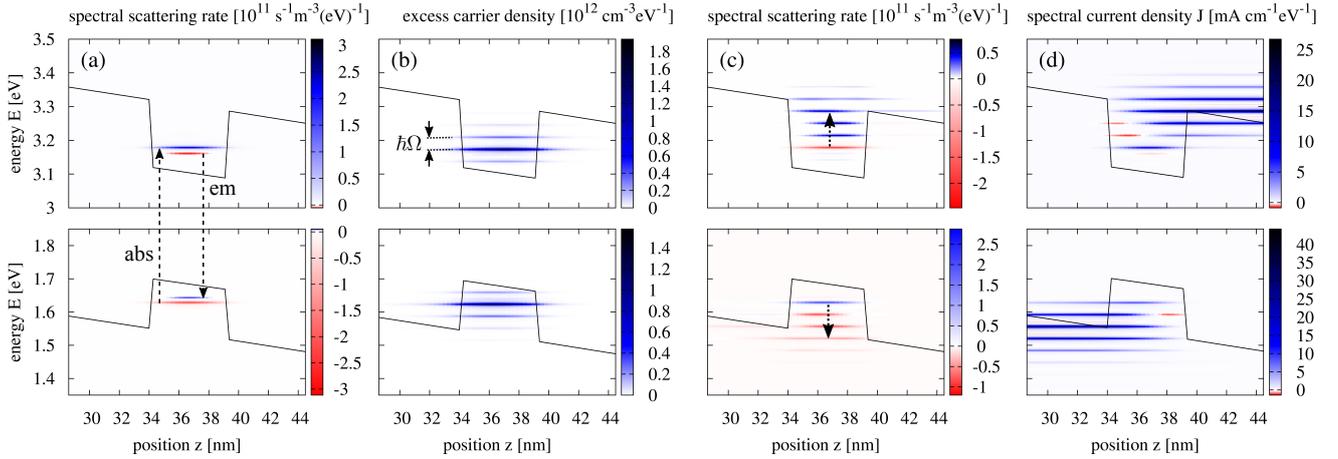} 
\caption{Spectral properties of  charge carriers in the quantum well region of the device from Fig.
\ref{fig:sqw_diode_structure} at contact Fermi-level separation of 1.3 V and  under illumination with monochromatic 
light at 1.55 eV and intensity of 17.7 kW/m$^2$: (a) Spatially resolved radiative generation and recombination
spectrum, (b) spectral excess carrier density with phonon satellites, (c) electron in-scattering rate due to coupling to
phonons, showing non-radiative charge transfer to higher states, (d) resulting current spectrum, with main contributions
well above the energy of photogeneration \cite{ae:spie10}.
\label{fig:spectral_sqw}}
\end{center}
\end{figure*}
The escape of charge carriers from quantum wells under illumination and at finite terminal bias
voltage is an excellent example of a quantum photovoltaic process that can only be described
properly by a quantum-kinetic approach, since neither the exact non-equilibrium carrier distribution
in the QW nor the phonon-assisted tunneling rate can be determined consistently in a semiclassical or ballistic
non-local approximation. The model system used for the investigation of the carrier escape process,
consisting of a $p$-$i$-$n$ diode of high band gap material with a single quantum well formed by a slab of low band 
gap material allocated at the center of the intrinsic region, is shown in Fig. \ref{fig:sqw_diode_structure}. The
quantum well thickness is 18 monolayers (ML), with the ML width given by half the lattice constant of the 
semiconductor material. The dopant concentration in the doped layers of 30 ML extension amounts to
$N_{d,a}=10^{18}$ cm$^{-3}$. The electronic structure is described by a simple two-band tight-binding model
\cite{boykin:96_1} providing band gaps $E_{g}^{b}=1.77$ eV and $E_{g}^{w}=1.42$ eV and band offsets $\Delta E_{C}=0.2$
eV, $\Delta E_{V}=0.15$ eV. The parameters for electron-photon and electron-phonon interactions correspond to the 
GaAs-Al$_{0.3}$Ga$_{0.7}$As material system (see Ref. \cite{ae:spie10} for details). Doping and band
offsets result in a band diagram in growth direction as shown schematically in Fig. \ref{fig:bandscheme}. At the operating point of the
solar cell, i.e. the device state where the product of current and voltage is maximum, the chemical potentials $\mu_{n,p}$ governing the injection of electrons and
holes at the respective contacts are split by the corresponding bias voltage, but the occupation of
the quantum well states is determined by the complex balance of the rates for photogeneration,
radiative recombination, carrier capture and carrier escape. Fig. \ref{fig:spectral_sqw} displays these rates
together with the resulting excess carrier concentration and current
spectrum at a contact Fermi-level splitting of 1.3 V and under illumination with monochromatic 
light at 1.55 eV and high intensity of 17.7 kW/m$^2$. The scattering rates are for electrons, a
positive value thus corresponds to inscattering of electrons in the conduction band and to 
outscattering of holes in the valence band. Due to the monochromatic illumination, the generation 
spectrum is narrow, but the interaction with phonons leads to a clear spectral separation of absorption and emission
lines on the one hand (a), and to the formation of phonon satellites in the spectral density of excess carriers on the
other hand (b). Current flow is not centered on the energy of generation, but on much higher energy,
as the wide triangular barriers prevent a direct tunneling escape from deep QW states (d). The
energy resolved phonon scattering rate (c) clearly evidences the phonon mediated scattering of carriers out of the low energy states where they are generated and into the less localized
higher energy states with enhanced tunneling escape rate. The NEGF formalism is thus able to provide
a complete microscopic picture of the thermionic emission process at arbitrary operating conditions.

\subsection{Generation and transport in Si-SiO$_{x}$ superlattice absorbers}
\begin{figure*}
\begin{center}
  (a)\includegraphics[width=6cm]{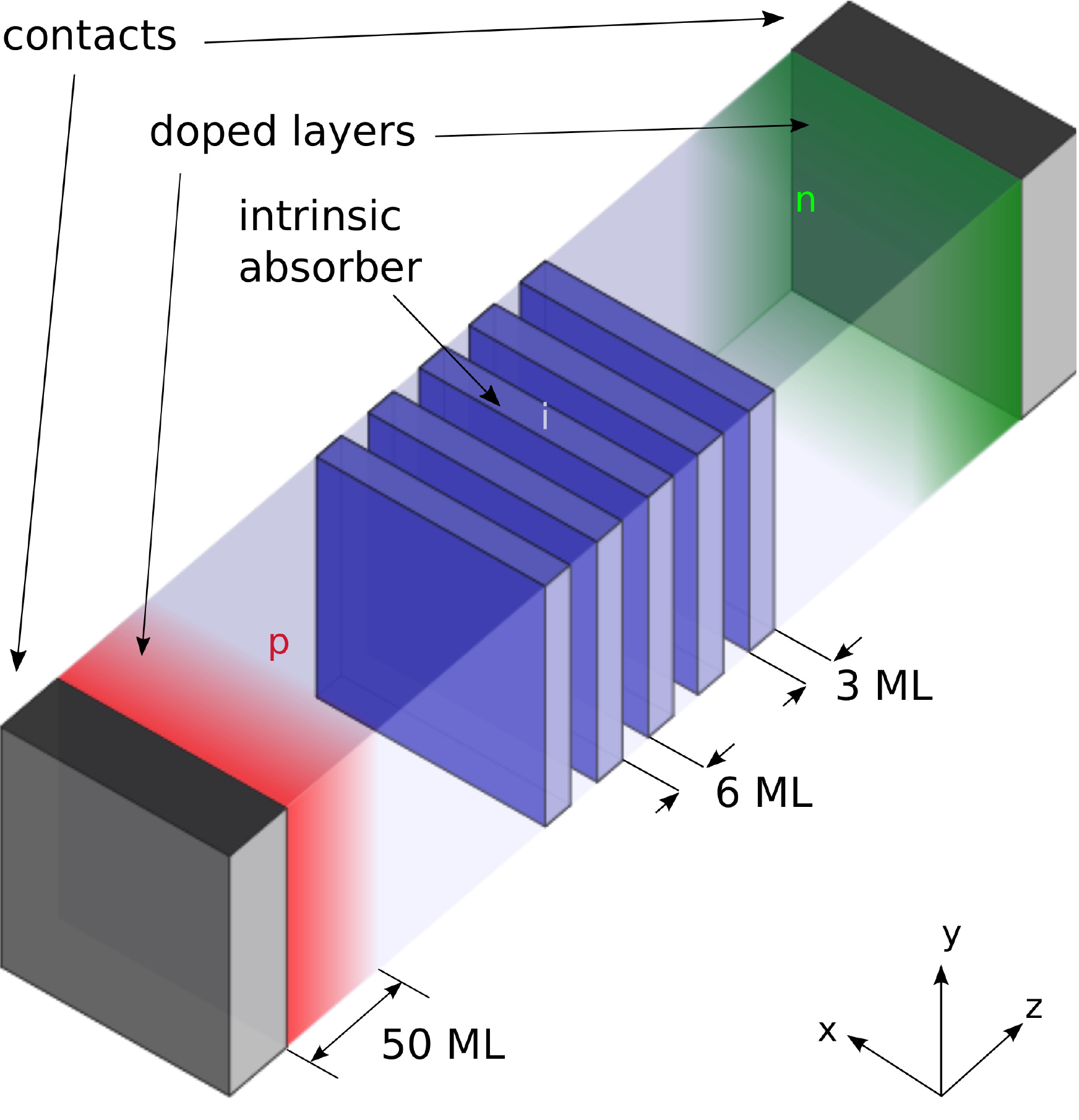}\qquad(b)\includegraphics[width=7cm]{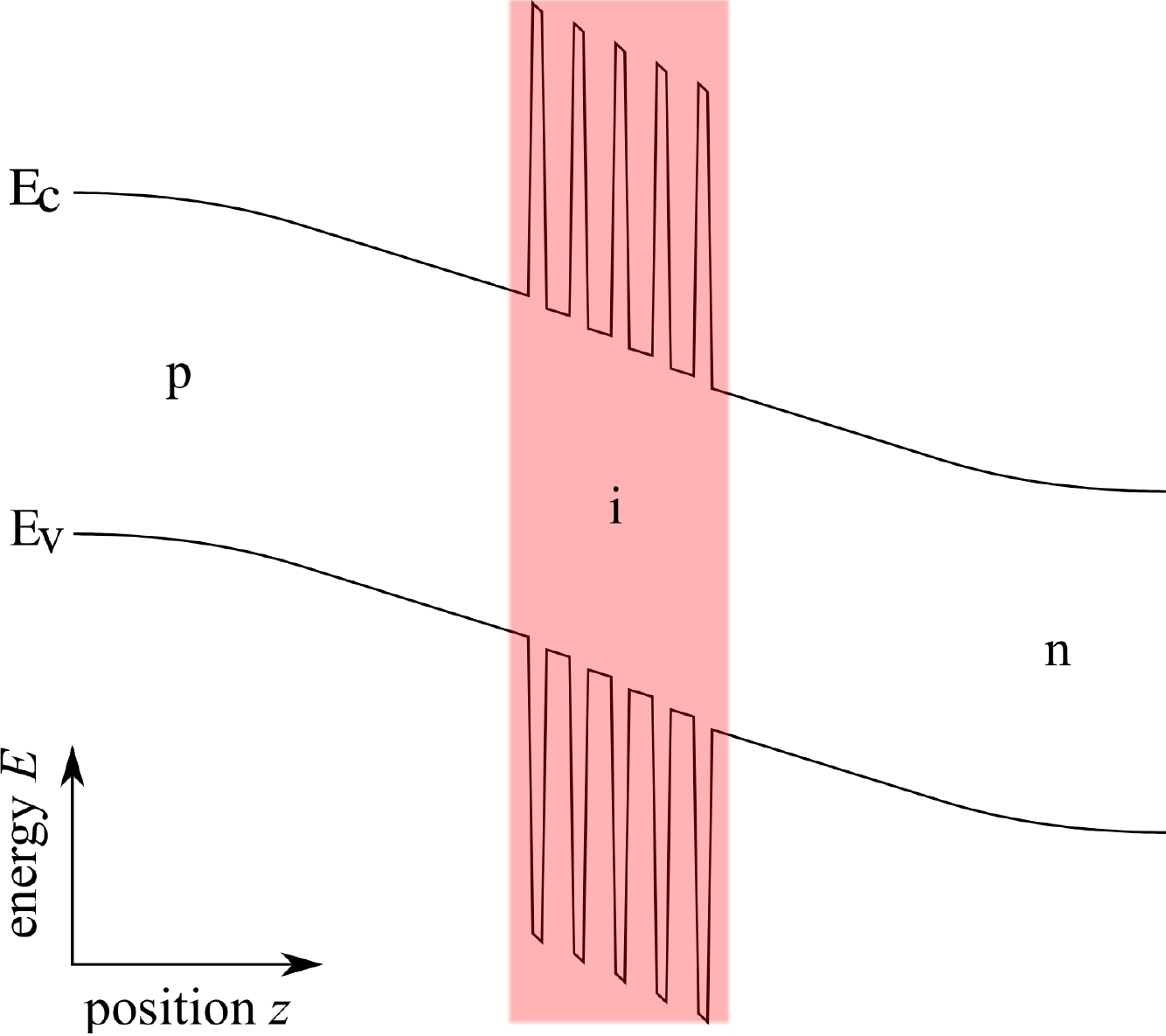}
  \caption{(a) Spatial structure and doping profile of the $p$-$i(SL)$-$n$ model system. The doping level is $N_{d}=10^{18}$ cm$^{-3}$ for
both electrons and holes, (b) band diagram of the $p$-$i(SL)$-$n$ model system
 with the active quantum well absorber region\label{fig:struct} \cite{ae:nrl_11}.}
  \end{center}
\end{figure*}
\begin{figure}[b!]
  \begin{center}
\includegraphics[width=7.1cm]{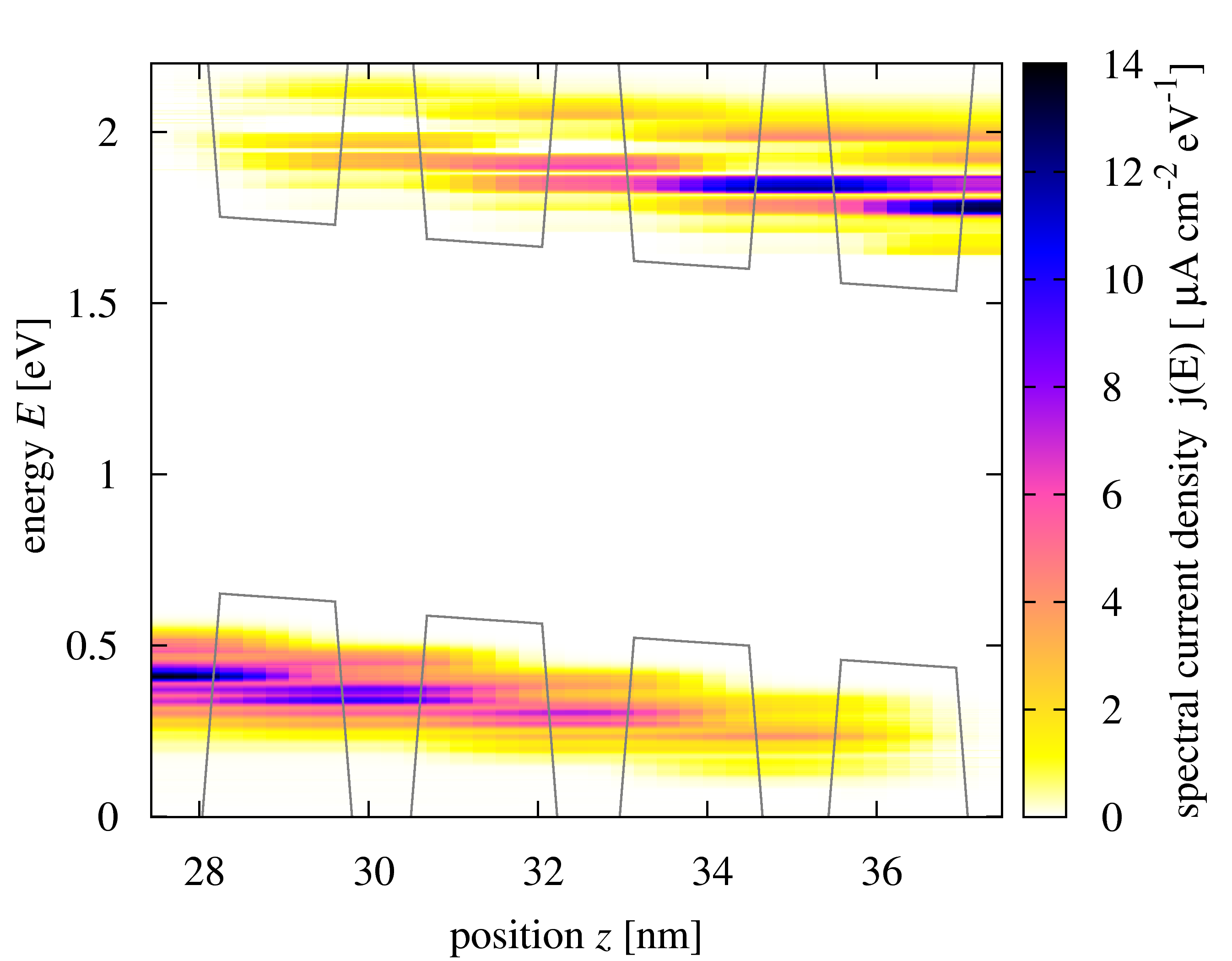}
   \caption{Spatially and energy resolved short-circuit photocurrent density in the quantum well region at short circuit
   conditions and under monochromatic illumination with energy $E_{\gamma}=1.65$ eV and intensity $I_{\gamma}=10$
   kW/m$^{2}$ \cite{ae:nrl_11}.\label{fig:pc_sl}} 
   \end{center} 
 \end{figure}

Owing to the large potential barriers for photoexcited charge carriers ($>$3 eV for both electrons
and holes), transport in Si-SiO$_{x}$ superlattice devices is restricted to quantum-confined
superlattice states. Due to the finite number of wells and large built-in fields, the electronic
spectrum can deviate considerably from the minibands of a regular superlattice. By including
the coupling of electrons to both photons and phonons, the NEGF theory is able to provide a microscopic picture of
indirect generation, carrier relaxation, and inter-well transport mechanisms beyond the ballistic
regime. The model system used to investigate generation and transport in indirect gap
superlattice absorbers is shown schematically in Fig.\,\ref{fig:struct}(a). It consists of a set of
four coupled quantum wells of 6 ML width with layers separated by oxide barriers of 3-ML
thickness, embedded in the intrinsic region of a Si $p$-$i$-$n$ diode. The thickness of both doped
and spacer layers is 50 ML. The monolayer thickness is half the Si lattice constant. For such thin oxide layers, the confinement and thus the effective
barrier height is reduced, and was chosen at 0.95 eV for both electron and holes, which is still 
large enough to prevent thermionic emission. The carriers are described using a multiband effective 
mass model, accounting for the different conduction band valleys as well
as light and heavy holes (details in \cite{ae:nrl_11}). The doping density is $N_{{\rm d}}=10^{18}$
cm$^{-3}$ for both electrons and holes. This composition and doping leads to the band diagram shown in Fig.\,\ref{fig:struct}(b).
In order to investigate the relaxation of photogenerated carriers, the superlattice region is
illuminated with a photon energy of 1.65 eV, thus exceeding the effective band gap, which at the
chosen barrier thickness lies around 1.3 eV. At this photon energy, both first and second
electronic minibands are populated, as can be seen in Fig. \label{fig:photrate_sl} displaying the spatially and
energetically resolved photogeneration rate at an intensity of 10 kW/m$^2$ and at vanishing bias
voltage. Remarkably, as shown in Fig. \ref{fig:pc_sl}, also the electronic current flows in both minibands,
relaxation due to scattering is thus not fast enough to confine transport to the band edge. It is
however obvious that transport of photocarriers across the superlattice is not ballistic, but
strongly affected by the inelastic interactions and best described by the sequential tunneling picture. 
\begin{figure}[b!]
  \begin{center}
   \includegraphics[width=7cm]{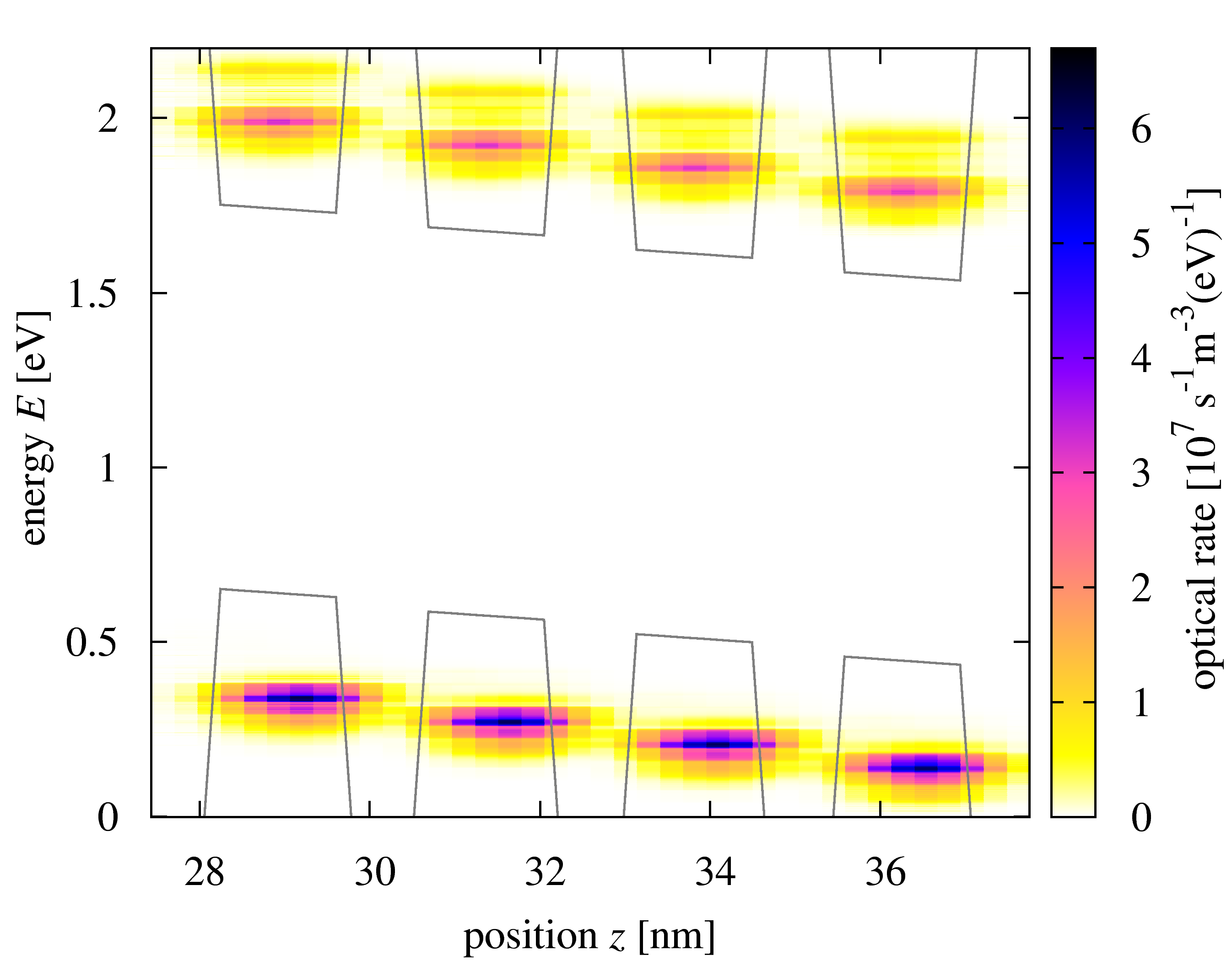}%\qquad\includegraphics[width=6cm]{figures/figure_6.png}
   \caption{Spatially and energy resolved charge carrier photogeneration rate in the quantum well region at short circuit conditions and
   under monochromatic illumination with energy $E_{\gamma}=1.65$ eV and intensity $I_{\gamma}=10$
   kW/m$^{2}$ \cite{ae:nrl_11}\label{fig:photrate_sl}}
   \end{center}
 \end{figure}
 
\begin{figure}[t!] 
 \begin{center}
  \includegraphics[width=8cm]{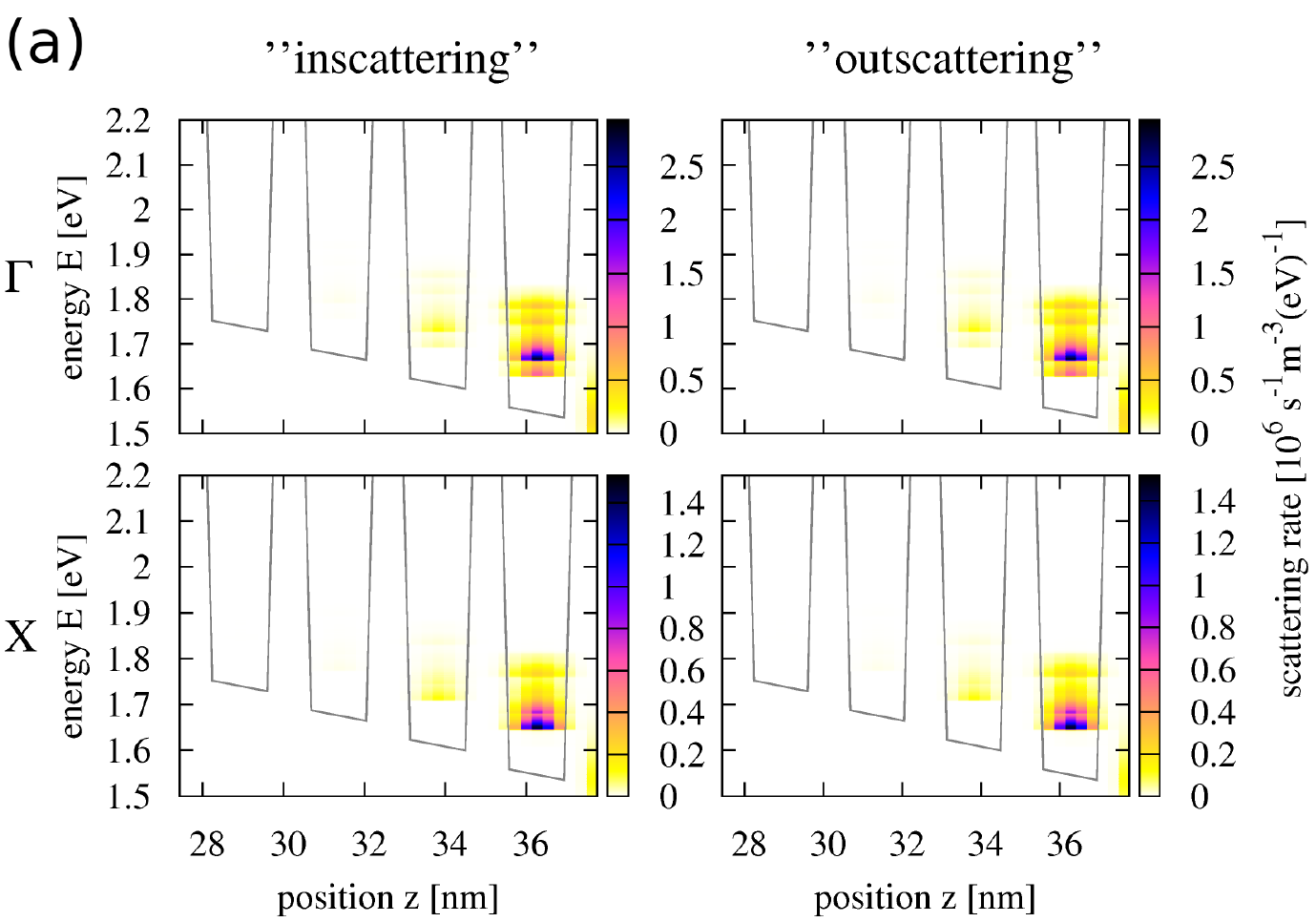}
  \includegraphics[width=8cm]{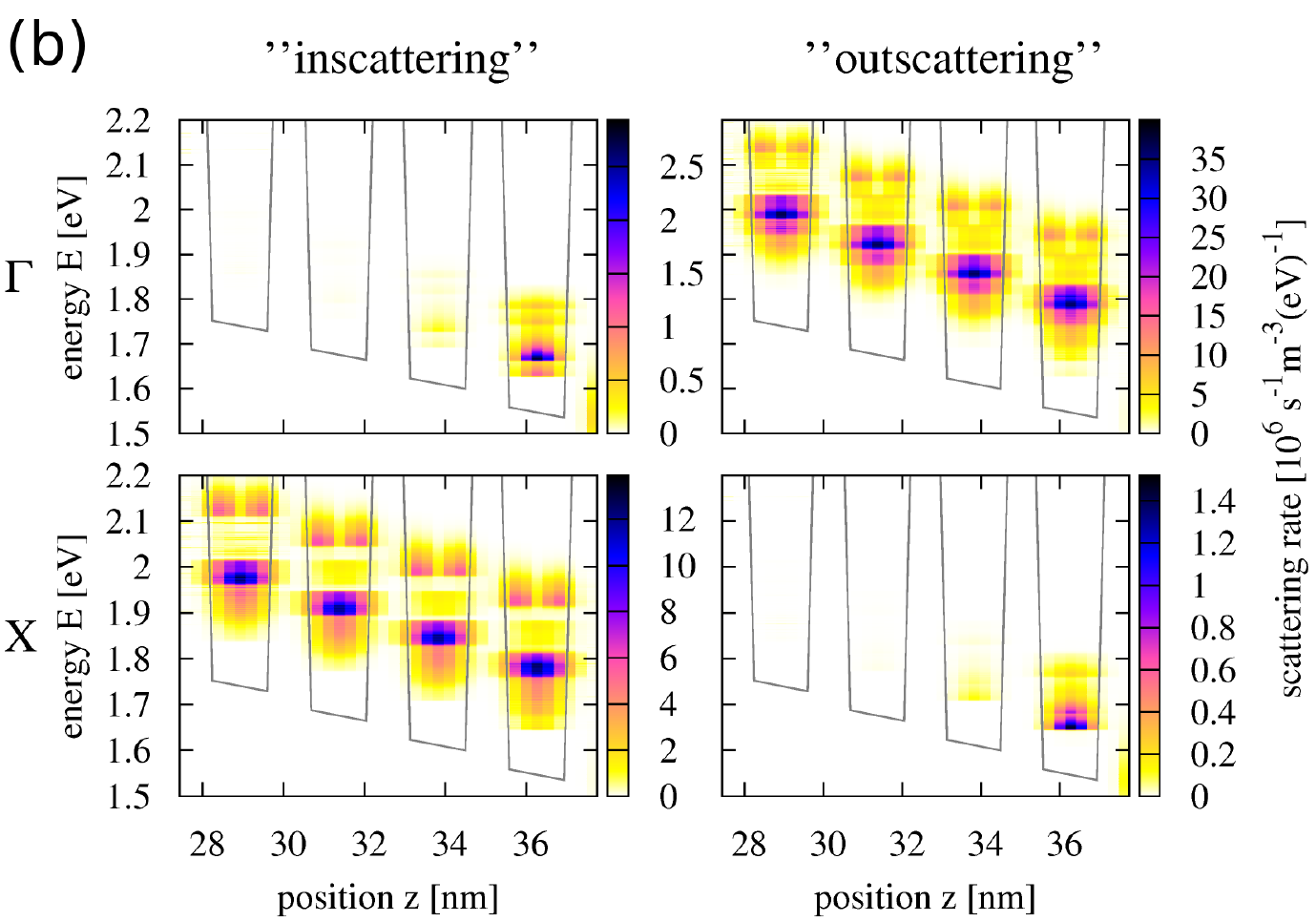}
  \caption{Spatially and energy resolved electron-phonon intervalley scattering rate at short
  circuit conditions (a) in the dark and (b) under monochromatic illumination with energy
  $E_{\gamma}=1.65$ eV and intensity $I_{\gamma}=10$ kW/m$^{2}$.\label{fig:phonrate_sl}}
  \end{center}
\end{figure}

It remains to consider the indirect photogeneration process. To this end, the phonon-mediated
intervalley in- and outscattering rates are displayed in Fig. \ref{fig:phonrate_sl} for both final electron state
in the Si valleys close to the X-point and the virtual intermediate state at the $\Gamma$-point.
While in-and out-scattering are balanced in the dark (a), there is a strong net charge
transfer from $\Gamma$ to $X$ under illumination, and the inscattering rate reflects the spectral 
pattern of the photogeneration.
\section{Conclusions}
The high efficiency concepts of the next generation of photovoltaic devices pose new requirements on
models for the  theoretical description of their optoelectronic characteristics. A theoretical framework suitable for this purpose
is available in the NEGF formalism, if the quantum optics formulation is unified with the picture of dissipative
quantum transport. The resulting theory is then able to provide insight into the modification of the central
physical processes of photogeneration, charge separation and carrier extraction due to the use of semiconductor nanostructures as absorber 
and conductor media. The general consideration all of the degrees of freedom on equal footing furthermore presents a suitable
starting point for consistent approximations in cases where some of the subsystems allow for a simplified treatment,
as e.g. for bulk modes, equilibrium distributions or negligible coupling parameters. While the
general framework is set, many challenges remain for the application to realistic photovoltaic
devices, both in advanced but fundamental aspects such as excitonic contributions to the optical
transitions and non-radiative recombination, as well as in the computational efficiency in treating
extended system without excessive loss of accuracy.

\begin{acknowledgements}
Financial support was provided by the German Federal Ministry of Education and Research (BMBF) under
Grant No. 03SF0352E.
\end{acknowledgements}

\bibliographystyle{spphys}       
%\bibliography{/home/aeberurs/Biblio/bib_files/negf,/home/aeberurs/Biblio/bib_files/qwsc,/home/aeberurs/Biblio/bib_files/bandstructure_TB,/home/aeberurs/Biblio/bib_files/pv,/home/aeberurs/Biblio/bib_files/aeberurs,/home/aeberurs/Biblio/bib_files/generation,/home/aeberurs/Biblio/bib_files/scqmoptics,/home/aeberurs/Biblio/bib_files/optical_modelling,/home/aeberurs/Biblio/bib_files/phonons,/home/aeberurs/Biblio/bib_files/sinova,/home/aeberurs/Biblio/bib_files/recombination}
%\bibliography{qpv_review}

\begin{thebibliography}{100}


\bibitem{yamaguchi:05}
M.~Yamaguchi, T.~Takamoto, K.~Araki, N.~Ekins-Daukes, Sol. Energy \textbf{79},
  78 (2005)

\bibitem{luque:97}
A.~Luque, A.~Mart\'i, Phys. Rev. Lett. \textbf{78}(26), 5014 (1997).

\bibitem{nozik:05}
A.J. Nozik, Inorganic Chemistry \textbf{44}(20), 6893 (2005).


\bibitem{ross:82}
R.T. Ross, A.J. Nozik, Journal of Applied Physics \textbf{53}(5), 3813 (1982).


\bibitem{rau:07}
U.~Rau, Phys. Rev. B \textbf{76}(8), 085303 (2007).

\bibitem{ned:99_2}
N.J. Ekins-Daukes, K.W.J. Barnham, J.P. Connolly, J.S. Roberts, J.C. Clark,
  G.~Hill, M.~Mazzer, Appl. Phys. Lett. \textbf{75}(26), 4195 (1999).

\bibitem{green:00}
M.A. Green, J. Mater. Sci. Eng. B \textbf{74}(1-3), 118  (2000)

\bibitem{berghoff:08}
B.~Berghoff, S.~Suckow, R.~Rolver, B.~Spangenberg, H.~Kurz, A.~Dimyati,
  J.~Mayer, Appl. Phys. Lett. \textbf{93}(13), 132111 (2008).

\bibitem{hermle:08}
M.~Hermle, G.~L\'etay, S.P. Philipps, A.W. Bett, Progress in Photovoltaics:
  Research and Applications \textbf{16}(5), 409 (2008).

\bibitem{martin:59}
P.C. Martin, J.~Schwinger, Phys. Rev. \textbf{115}(6), 1342 (1959).

\bibitem{schwinger:61}
J.~Schwinger, J. Math. Phys. \textbf{2}, 407 (1961)

\bibitem{kadanoff:62} 
L.P. Kadanoff, G.~Baym, \emph{Quantum Statistical Mechanics} (Benjamin,
  Reading, Mass., 1962)

\bibitem{keldysh:65}
L.~Keldysh, Sov. Phys.-JETP. \textbf{20}, 1018 (1965)

\bibitem{langreth:76}
D.~Langreth, in \emph{Linear and Non-linear Electron Transport in solids}
  \textbf{17}, 3 (1976)

\bibitem{fetter_walecka}
A.L. Fetter, J.D. Walecka, \emph{Quantum theory of many-particle systems}
  (McGraw-Hill, San Francisco, 1971)

\bibitem{lake:97}
R.~Lake, G.~Klimeck, R.~Bowen, D.~Jovanovic, J. Appl. Phys. \textbf{81}, 7845
  (1997)

\bibitem{luisier:06_2}
M.~Luisier, A.~Schenk, W.~Fichtner, G.~Klimeck, Phys. Rev. B \textbf{74}(20),
  205323 (2006).

\bibitem{jin:06_2}
S.~Jin, Y.J. Park, H.S. Min, J. Appl. Phys. \textbf{99}, 123719 (2006)

\bibitem{henrickson:94}
L.E. Henrickson, A.J. Glick, G.W. Bryant, D.F. Barbe, Phys. Rev. B
  \textbf{50}(7), 4482 (1994).

\bibitem{tian:98}
W.~Tian, S.~Datta, S.~Hong, R.~Reifenberger, J.I. Henderson, C.P. Kubiak, J.
  Chem. Phys. \textbf{109}(7), 2874 (1998).

\bibitem{xue:01}
Y.~Xue, S.~Datta, M.A. Ratner, J. Chem. Phys. \textbf{115}(9), 4292 (2001).

\bibitem{brandbyge:02}
M.~Brandbyge, J.L. Mozos, P.~Ordej\'on, J.~Taylor, K.~Stokbro, Phys. Rev. B
  \textbf{65}(16), 165401 (2002).

\bibitem{dicarlo:05}
A.~Di~Carlo, A.~Pecchia, L.~Latessa, T.~Frauenheim, G.~Seifert, in
  \emph{Lecture Notes in Physics}, vol. 680 (Springer, 2005), p. 153

\bibitem{stokbro:05}
K.~Stokbro, J.~Taylor, M.~Brandbyge, H.~Guo, in \emph{Lecture Notes in Physics}
  (Springer, 2005)

\bibitem{rocha:06}
A.R. Rocha, V.M. Garc\'{\i}a-Su\'{a}rez, S.~Bailey, C.~Lambert, J.~Ferrer,
  S.~Sanvito, Phys. Rev. B \textbf{73}(8), 085414 (2006).

\bibitem{thygesen:08}
K.S. Thygesen, A.~Rubio, Phys. Rev. B \textbf{77}(11), 115333 (2008).

\bibitem{lake:92}
R.~Lake, S.~Datta, Phys. Rev. B \textbf{46}(8), 4757 (1992).

\bibitem{frederiksen:04}
T.~Frederiksen, Inelastic electron transport in nanosystems.
\newblock Master's thesis, Technical University of Denmark (2004)

\bibitem{pecchia:07}
A.~Pecchia, G.~Romano, A.D. Carlo, Phys. Rev. B \textbf{75}(3), 035401 (2007).

\bibitem{lu:07}
J.T. L\"u, J.S. Wang, Phys. Rev. B \textbf{76}(16), 165418 (2007).

\bibitem{wang:07}
J.S. Wang, N.~Zeng, J.~Wang, C.K. Gan, Phys. Rev. E \textbf{75}(6), 061128
  (2007).
  
\bibitem{groshev:91}
A.~Groshev, T.~Ivanov, V.~Valtchinov, Phys. Rev. Lett. \textbf{66}(8), 1082
  (1991).


\bibitem{chen:91_2}
L.Y. Chen, C.S. Ting, Phys. Rev. B \textbf{44}(11), 5916 (1991).

\bibitem{hershfield:91}
S.~Hershfield, J.H. Davies, J.W. Wilkins, Phys. Rev. Lett. \textbf{67}(26),
  3720 (1991).

\bibitem{meir:93}
Y.~Meir, N.S. Wingreen, P.A. Lee, Phys. Rev. Lett. \textbf{70}(17), 2601
  (1993).

\bibitem{binder:95}
R.~Binder, S.W. Koch, Prog. Quantum Electron. \textbf{19}, 307 (1995)

\bibitem{haug:96}
H.~Haug, A.P. Jauho, \emph{Quantum kinetics in transport and optics of
  semiconductors} (Springer, Berlin, 1996)

\bibitem{pereira:98}
M.F. Pereira, K.~Henneberger, Phys. Rev. B \textbf{58}, 2064 (1998)

\bibitem{hannewald:01}
K.~Hannewald, Femtosecond dynamics of absorption and luminescence in optically
  excited semiconductors: theory and simulation.
\newblock Ph.D. thesis, Jena (2001)

\bibitem{schaefer:02}
W.~Sch\"afer, M.~Wegener, \emph{Semiconductor Optics and Transport Phenomena}
  (Springer, Berlin, 2002)

\bibitem{haug:04}
H.~Haug, S.W. Koch, \emph{Quantum Theory of the Optical and Electronic
  Properties of Semiconductors} (World Scientific, 2004)

\bibitem{caroli:71}
C.~Caroli, R.~Combescot, P.~Nozi\`{e}res, D.~Saint-James, J. Phys. C: Solid
  State Phys. \textbf{4}, 916 (1971)

\bibitem{svizhenko:02}
A.~Svizhenko, M.P. Anantram, T.R. Govindan, B.~Biegel, R.~Venugopal, J. Appl.
  Phys. \textbf{91}(4), 2343 (2002).

\bibitem{guo:02}
J.~Guo, M.S. Lundstrom, IEEE Trans. Electron Devices \textbf{49}, 1897 (2002)

\bibitem{ren:03}
Z.~Ren, R.~Venugopal, S.~Goasguen, S.~Datta, IEEE, M.S. Lundstrom, IEEE Trans.
  Electron Devices \textbf{50}, 1914 (2003)

\bibitem{jin:06}
S.~Jin, {M}odeling of {Q}uantum {T}ransport in {N}ano-{S}cale {MOSFET}
  {D}evices.
\newblock Ph.D. thesis, Seoul National University (2006)

\bibitem{martinez:07}
A.~Martinez, M.~Bescond, J.~Barker, A.~Svizhenko, M.~Anantram, C.~Millar,
  A.~Asenov, IEEE Trans. Electron Devices \textbf{54}(9), 2213 (2007).
  
\bibitem{guo:05}
J.~Guo, J. Appl. Phys. \textbf{98}(6), 063519 (2005).

\bibitem{pourfath:06}
M.~Pourfath, H.~Kosina, S.~Selberherr, J. Phys: Conf. Ser. \textbf{38}, 29
  (2006)

\bibitem{kim:88}
G.~Kim, G.B. Arnold, Phys. Rev. B \textbf{38}(5), 3252 (1988).

\bibitem{anda:91}
E.V. Anda, F.~Flores, J. Phys.: Condens. Matter. \textbf{3}, 9087 (1991)

\bibitem{kim:95}
G.~Kim, H.H. Suh, E.H. Lee, Phys. Rev. B \textbf{52}(4), 2632 (1995).

\bibitem{ogawa:99}
M.~Ogawa, T.~Sugano, R.~Tominaga, T.~Miyoshi, Physica B \textbf{272}, 167
  (1999)

\bibitem{ogawa:00}
M.~Ogawa, T.~Sugano, T.~Miyoshi, Solid-State Electron. \textbf{44}, 1939 (2000)

\bibitem{rivas:01}
C.~Rivas, R.~Lake, G.~Klimeck, W.R. Frensley, M.V. Fischetti, P.E. Thompson,
  S.L. Rommel, P.R. Berger, Appl. Phys. Lett. \textbf{78}(6), 814 (2001).
  
\bibitem{rivas:03}
C.~Rivas, R.~Lake, W.R. Frensley, G.~Klimeck, P.E. Thompson, K.D. Hobart, S.L.
  Rommel, P.R. Berger, J. Appl. Phys. \textbf{94}(8), 5005 (2003).

\bibitem{luisier:10}
M.~Luisier, G.~Klimeck, Journal of Applied Physics \textbf{107}(8), 084507
  (2010).

\bibitem{lee_prb:02}
S.C. Lee, A.~Wacker, Phys. Rev. B \textbf{66}, 245314 (2002)

\bibitem{kubis:09}
T.~Kubis, C.~Yeh, P.~Vogl, A.~Benz, G.~Fasching, C.~Deutsch, Phys. Rev. B
  \textbf{79}, 195323 (2009)

\bibitem{henrickson:02}
L.E. Henrickson, J. Appl. Phys \textbf{91}, 6273 (2002)

\bibitem{stewart:04}
D.A. Stewart, F.~Leonard, Phys. Rev. Lett. \textbf{93}, 107401 (2004)

\bibitem{stewart:05}
D.A. Stewart, F.~Leonard, Nano Lett. \textbf{5}, 219 (2005)

\bibitem{guo:06}
J.~Guo, M.A. Alam, Y.~Yoon, Appl. Phys. Lett. \textbf{88}(13), 133111 (2006).

\bibitem{steiger:iwce_09}
S.~Steiger, R.G. Veprek, B.~Witzigmann, in \emph{Proceedings - 2009 13th
  International Workshop on Computational Electronics, IWCE 2009} (2009)

\bibitem{evers:04}
F.~Evers, F.~Weigend, M.~Koentopp, Phys. Rev. B \textbf{69}(23), 235411 (2004).

\bibitem{ae:prb_08}
U.~Aeberhard, R.H. Morf, Phys. Rev. B \textbf{77}, 125343 (2008)

\bibitem{ae:thesis}
U.~Aeberhard, A {M}icroscopic {T}heory of {Q}uantum {W}ell {P}hotovoltaics.
\newblock Ph.D. thesis, ETH Zurich (2008)

\bibitem{wuerfel:book_05}
P.~W\"urfel, \emph{Physics of Solar Cells} (Wiley-VCH, 2005)

\bibitem{danielewicz:84}
P.~Danielewicz, Ann. Phys. \textbf{152}, 239 (1984)

\bibitem{rammer:86}
J.~Rammer, H.~Smith, Rev. Mod. Phys. \textbf{58}, 323 (1986)

\bibitem{datta:95}
S.~Datta, \emph{Electronic Transport in Mesoscopic Systems} (Cambridge
  University Press, 1995)

\bibitem{henneberger:88_3}
K.~Henneberger, H.~Haug, Phys. Rev. B \textbf{38}, 9759 (1988)

\bibitem{hedin:65}
L.~Hedin, Phys. Rev. \textbf{139}(3A), A796 (1965).

\bibitem{mozyrsky:07}
D.~Mozyrsky, I.~Martin, Optics Communications \textbf{277}, 109 (2007)

\bibitem{jahnke:95}
F.~Jahnke, S.W. Koch, Phys. Rev. A \textbf{52}(2), 1712 (1995).

\bibitem{srivastava:90}
G.~Srivastava, \emph{The Physics of phonons} (IOP Publishing, Bristol, England,
  A. Hilger, 1990)

\bibitem{luisier:09}
M.~Luisier, G.~Klimeck, Physical Review B (Condensed Matter and Materials
  Physics) \textbf{80}(15), 155430 (2009).

\bibitem{guinea:83}
F.~Guinea, C.~Tejedor, F.~Flores, E.~Louis, Phys. Rev. B \textbf{28}(8), 4397
  (1983).

\bibitem{lopez_sancho:84}
M.P.L. Sancho, J.M.L. Sancho, J.~Rubio, J. Phys. F: Met. Phys. \textbf{14}(5),
  1205 (1984).

\bibitem{umerski:97}
A.~Umerski, Phys. Rev. B \textbf{55}(8), 5266 (1997).

\bibitem{lee:81_2}
D.H. Lee, J.D. Joannopoulos, Phys. Rev. B \textbf{23}(10), 4997 (1981).

\bibitem{chang:82}
Y.C. Chang, J.~Schulman, Phys. Rev. B \textbf{25}, 3975 (1982)

\bibitem{bouwen:95}
R.~Bouwen, W.~Frensley, G.~Klimeck, R.~Lake, Phys. Rev. B \textbf{52}, 2754
  (1995)

\bibitem{rahachou:05}
A.I. Rahachou, I.V. Zozoulenko, Phys. Rev. B \textbf{72}(15), 155117 (2005).

\bibitem{velasco:88}
V.R. Velasco, F.~Garcia-Moliner, L.~Miglio, L.~Colombo, Phys. Rev. B
  \textbf{38}(5), 3172 (1988).

\bibitem{garcia_moliner:86}
F.~Garcia-Moliner, V.R. Velasco, Progress in Surface Science \textbf{21}(2), 93
   (1986).

\bibitem{mahan:90}
G.D. Mahan, \emph{Many-Particle Physics, 2nd ed.} (Plenum, New York, 1990)

\bibitem{hyldgaard:94}
P.~Hyldgaard, S.~Hershfield, J.~Davies, J.~Wilkins, Ann. Phys. \textbf{236}, 1
  (1994)

\bibitem{haug:88}
H.~Haug, K.~Henneberger, Phys. Rev. B \textbf{38}, 9759 (1988)

\bibitem{henneberger:96}
K.~Henneberger, S.W. Koch, Phys. Rev. Lett. \textbf{76}(11), 1820 (1996).

\bibitem{richter:08}
F.~Richter, M.~Florian, K.~Henneberger, Phys. Rev. B \textbf{78}(20), 205114
  (2008).

\bibitem{meir:92}
Y.~Meir, N.~Wingreen, Phys. Rev. Lett. \textbf{68}, 2512 (1992)

\bibitem{henneberger:09}
K.~Henneberger, Phys. Status Solidi B \textbf{246}, 283 (2009)

\bibitem{wuerfel:82}
P.~W\"urfel, J. Phys. C: Solid State Phys. \textbf{15}, 3967 (1982)

\bibitem{ae:prb_11}
U.~Aeberhard, unpublished, arXiv/1012.5462

\bibitem{shockley:52}
W.~Shockley, W.T. Read, Phys. Rev. \textbf{87}, 835 (1952)

\bibitem{hall:52}
R.~Hall, Phys. Rev. \textbf{87}, 387 (1952) 

\bibitem{a_d_maur:08}
M.~Auf~der Maur, M.~Povolotskyi, F.~Sacconi, A.~Pecchia, G.~Romano, G.~Penazzi,
  A.~Di~Carlo, Optical and Quantum Electronics \textbf{40}, 1077 (2008).

\bibitem{wacker:02}
A.~Wacker, Phys. Rep. \textbf{357}, 1 (2002)

\bibitem{pourfath:09}
M.~Pourfath, H.~Kosina, Journal of Computational Electronics  (2009)

\bibitem{kubis:06_2}
T.~Kubis, P.~Vogl, J. Comp. Electron. \textbf{6}, 183 (2007) 

\bibitem{steiger:thesis}
S.~Steiger, Modeling {N}ano-{LED}.
\newblock Ph.D. thesis, ETH Zurich (2009)

\bibitem{ae:spie10}
U.~Aeberhard,  (SPIE, 2010), vol. 7597, p. 759702.

\bibitem{ae:solmat_10}
U.~Aeberhard, Sol. Energy Mater. Sol. Cells \textbf{94}, 1897 (2010) 

\bibitem{ae:nrl_11}
U.~Aeberhard, Nanoscale Res. Lett. \textbf{6}, 242 (2011).

\bibitem{boykin:96_1}
T.B. Boykin, Phys. Rev. B \textbf{54}, 7670 (1996)

\end{thebibliography}

\end{document}